# Andreev Reflection in the Fractional Quantum Hall State

Önder Gül,[1,][*] Yuval Ronen,[1,][*] Si Young Lee,[1,2,][*] Hassan Shapourian,[1,3] Jonathan Zauberman,[1] Young Hee Lee,[2,4] Kenji Watanabe,[5] Takashi Taniguchi,[6] Ashvin Vishwanath,[1] Amir Yacoby,[1] and Philip Kim[1,][✉]

[1]Department of Physics, Harvard University, Cambridge, MA, USA
[2]Center for Integrated Nanostructure Physics (CINAP), Institute for Basic Science (IBS), Suwon, Republic of Korea
[3]Department of Physics, Massachusetts Institute of Technology, Cambridge, MA, USA
[4]Department of Energy Science, Sungkyunkwan University, Suwon, Republic of Korea
[5]Research Center for Functional Materials, National Institute for Materials Science, Tsukuba, Japan
[6]International Center for Materials Nanoarchitectonics, National Institute for Materials Science, Tsukuba, Japan

We construct high-quality graphene-based van der Waals devices with narrow superconducting niobium nitride (NbN) electrodes, in which superconductivity and robust fqH coexist. We find crossed Andreev reflection (CAR) across the superconductor separating two fqH edges. Our observed CAR probabilities in the particle-like fractional fillings are markedly higher than those in the integer and hole-conjugate fractional fillings and depend strongly on temperature and magnetic field unlike the other fillings. Further, we find a filling-independent CAR probability in integer fillings, which we attribute to spin-orbit coupling in NbN allowing for Andreev reflection between spin-polarized edges. These results provide a route to realize novel topological superconducting phases in fqH–superconductor hybrid devices based on graphene and NbN.

## I. INTRODUCTION

Topological superconductors are predicted to represent a phase of matter with nonlocal properties, providing a robustness suitable for quantum computing[1–5]. A theoretical proposal to synthesize a topological superconductor from a topological insulator and a conventional (s-wave) superconductor has motivated hybrid approaches to realize Majorana modes. Besides topological insulators[6–8], these approaches now include spin-orbit coupled semiconductors[9–14], magnetic atom chains[15], and integer quantum Hall edges[16–18]—all in combination with a superconductor—offering either a testbed for or a route towards topological qubits. Common to all of these is the noninteracting description of charge carriers and Ising topological order which is insufficient for universal quantum computation[4]. These approaches, however, can be extended to the computationally universal Fibonacci order[19] predicted to emerge in a coupled parafermion array[20].

Parafermions, unlike Majoranas, require electron-electron interactions to form, which result in richer non-Abelian braiding statistics[21]. An established condensed matter system that forms with interactions is the fqH state, which is the basis of different approaches for synthesizing parafermions[19–28]. The primary approach—combining fqH, appearing in semiconductor heterostructures, with superconductivity[20,23–27]—has so far presented two major experimental challenges. First, the strong magnetic fields required for fqH suppress superconductivity[16–18,29]. Second, coupling a superconductor to a semiconductor heterostructure[30] can be difficult, often leading to a nontransparent interface. Here, we overcome these challenges by using graphene-based van der Waals (vdW) heterostructures coupled to superconducting niobium nitride (NbN). The high device quality decreases the magnetic fields required for robust fqH to the regime where NbN remains superconducting owing to its large critical field.

---
*These authors contributed equally.
✉Corresponding author.
philipkim@g.harvard.edu

The superconductor edge-contact to graphene provides an interface transparent enough to allow Andreev reflection in quantum Hall edges.

## II. THE DEVICE AND CROSSED ANDREEV REFLECTION IN FQH

Figure 1a shows the schematic of our vdW heterostructure, consisting of single-layer graphene as the conducting channel, which is first encapsulated by hexagonal boron nitride dielectric and then by graphite on both top and bottom. This heterostructure maximizes the channel mobility owing to the metallic graphite layers screening remote impurities[31,32], which is essential for reaching the fqH phase at magnetic fields low enough to allow superconductivity. Figure 1b shows a typical device, including the heterostructure (purple) and a <100 nm wide NbN superconductor (blue) which is sufficiently narrow (coherence length ∼50 nm)[17] to ensure crossed Andreev reflection (CAR) between the quantum Hall edges on both sides (Figure 6), a necessary ingredient for realizing parafermions[33]. CAR dramatically affects transport: when the injected electron-like charges are drained from the superconductor, hole-like charges propagate away (Figure 1c).

We have measured the resistance $R_{CAR}=V_{CAR}/I_{exc}$ as well as the Hall resistance $R_{XY}=V_{XY}/I_{exc}$ as a function of gate voltage (charge carrier density) at a magnetic field $B=14$ T for different temperatures $T$ (Figure 1d). Here, $V_{CAR}$ is the potential of the edge mode propagating away from the grounded superconductor, $V_{XY}$ the Hall voltage, and $I_{exc}$ the bias current (see Figure 1b for the circuit). At low $T$, $R_{CAR}$ becomes negative for quantized values of $R_{XY}$. We find an $R_{CAR}<0$ for both integer fillings 1 and 2[17], and importantly for several fractional fillings 1/3, 2/5, 2/3, 5/3—our main finding (blow-up shown in Figure 1e). An $R_{CAR}<0$ indicates that the electron-like carriers drained from the superconductor produce hole-like carriers with opposite charge, a direct result of crossed Andreev reflection, which reverses the sign of the edge potential. $R_{CAR}$ acquires positive values either when $R_{XY}$ is nonquantized and the bulk of the device conducts, or when superconductivity is suppressed with increasing $T$—both destroying CAR as expected. We confirm that our narrow NbN superconducts at 14 T for $T<8$ K by measuring a strip with identical dimensions as the one coupled to the quantum Hall edges (Figure 1d inset). In a separate cooldown with $T$ reaching 15 mK, we find $R_{CAR}<0$ for several fractional fillings for a wide range of $B$ (Figure 1f). At this low temperature, CAR in the fqH state can be observed in magnetic fields as low as 3 T (filling 2/3).

### III. SPIN-ORBIT COUPLING

Figure 2a shows $R_{CAR}$ with the accompanying longitudinal resistance $R_{XX}$ as a function of filling $\nu$. For all integer fillings $R_{XX}$, which measures bulk conduction, is much smaller in amplitude than $R_{CAR}$, linking $R_{CAR}$ strictly to the potential of the edge mode leaving the superconductor. We find a negative edge potential with consistently increasing amplitude for lower fillings, which remains negative for all integer $\nu$ and measured $B$ demonstrating the robustness of CAR (Figure 2a inset).

The increasing amplitude of $R_{CAR}$ with decreasing $\nu$ is connected to the decreasing number of edge modes, which increases $R_{XY}$. This dependence of $R_{CAR}$ on $\nu$ can be understood by introducing $p_{CAR}=-V_{CAR}/V$, the probability of crossed Andreev reflection, where $V$ is the potential of the incoming edge mode. The Hall voltage constrains $V_{XY}=V-V_{CAR}$, leading to the proportional relation $R_{CAR}=-R_{XY}/(1+p_{CAR}^{-1})$. This allows us to calculate $p_{CAR}$ from the measured $R_{CAR}$, and thus to directly compare the CAR rate between different fillings. Figure 2b shows that $p_{CAR}$ is comparable for all integer $\nu$, including the spin-polarized $\nu=1$. This striking finding provides evidence for the presence of strong spin-orbit coupling (SOC) in the NbN superconductor. Without this SOC, CAR cannot occur in the spin-polarized edges due to the s-wave superconducting pairing in NbN. Our Bogoliubov-de Gennes spectrum, calculated for integer $1\leq\nu\leq6$, supports this observation (Figure 7; see Figure 2b inset and Appendix B for the theory model). Figure 2c and d show the energy spectrum of $\nu=1$ and 2 without and with SOC, represented in our Hamiltonian by the term $\lambda_R$. A pairing gap $\Delta_{ind}$ does not open without SOC (Figure 2c). The inclusion of SOC allows for pairing ($\Delta_{ind}>0$) between the two counterpropagating spin-up edges of $\nu=1$, and separately between the two additional counterpropagating spin-down edges of $\nu=2$ (Figure 2d)—pairing between the edges with opposite spin polarization is forbidden owing to the difference in their Fermi wave vector $k_F$. Interestingly, our comparable $p_{CAR}$ for all integer $\nu$ and its magnetic field insensitivity for small $I_{exc}$ and $T$ (compared to $\Delta_{ind}$) is consistent with a topologically nontrivial $\Delta_{ind}$.

Our $p_{CAR}$ remains much smaller than unity[34] accounted for in our model by a parameter $Z$ to represent various processes which lead to the tunneling of incoming charges to outgoing edge modes without Andreev reflection (Figure 2b inset), analogous to that in Blonder-Tinkham-Klapwijk formalism[35]. Cotunneling and subgap transport due to vortices are the leading mechanisms contributing to $Z$ expected for our type-II superconductor. From the measured $p_{CAR}$ we extract $Z\sim0.45$ which is constant for all integer $\nu$. A small $p_{CAR}$ is microscopically related to disorder in the graphene–superconductor interface, which we effectively model with disordered pairing[36] (see CAR resistance in Appendix B2, Figure 25 and 26). Our model does not include edge reconstruction[37] or charge accumulation at the superconductor interface (both discussed below) but accounts for Andreev edge states demonstrated in quantum Hall–superconductor hybrids without counterpropagating edge modes[16,18,29,38,39]. Andreev edge states do not result in a pairing gap (Figure 20) and experimentally produce a response sensitive to small changes in gate voltage, magnetic field and the length of the superconductor interface, which averages to zero[18]. In contrast, our CAR response is robust, consistently negative, and independent of the interface length (Figure 6), suggesting an induced pairing.

A potential link between CAR and the induced pairing may also be revealed by spectroscopy. We have varied the energy of the injected charges by tuning $V$, serving as the bias voltage (Figure 1b), and found CAR to be limited to low energies (Figure 2e and f). Notably, the energy range of CAR in $\nu=2$ is larger than in $\nu=6$, consistent with a larger pairing gap for lower fillings owing to their edge modes' closer proximity to the superconductor (Figure 26). This gate dependence excludes, for integer fillings, the effects of nonproximitized residual edge modes originating from the work function mismatch between the graphene channel and the superconductor, which would result in $\nu$-independent spectroscopic properties. (A work function mismatch tends to produce a heavily doped region near the contact[40], whose precise physical extent in our device cannot be measured directly.)

### IV. MAGNETIC FIELD DEPENDENCE

We now extend our analysis of $p_{CAR}$ to fractional $\nu$ and a larger $B$ range. Figure 3a shows $p_{CAR}$ and the accompanying $\nu\cdot R_{XX}$ ($R_{XX}$ normalized for different $\nu$), with the measured $R_{CAR}$ and $R_{XX}$ at $B=9$ T plotted in Figure 3b. For all fractional $\nu$ with $R_{CAR}<0$, we find a negligibly small $R_{XX}$ which excludes bulk conduction. Bulk conduction does, however, result in $R_{CAR}>0$ observed for lower values of $B$ or for fractional $\nu$ with smaller excitation gaps. Figure 3c shows the $B$ dependence of $p_{CAR}$ for several fractional and integer $\nu$ along with the expected $p_{CAR}$ for our $Z$. In this $B$ range, we again find for several fillings a $p_{CAR}$ that does not depend on $B$ and is comparable between different $\nu$, this time including several fractional $\nu$ (Figure 3b inset). Importantly, this observation is contrasted by particle-like fillings $\nu=1/3$ and $4/3$, which present well-developed fqH (negligible $R_{XX}$) together with a strong magnetic field dependence. This behavior can be a result of edge reconstruction, intrinsic to the superconducting pairing of fqH edges, or due to a $B$-dependent $Z$ for certain fractional fillings. Identifying these scenarios theoretically is challenging due to the interacting many-body nature of fqH not allowing for our Bogoliubov-Dirac-Landau analysis. Experimentally, we find a $p_{CAR}$ that strongly depends on $B$ (and $T$, presented below) only for the particle-like fqH states whose $p_{CAR}$ is significantly larger than that for integer $\nu$. Such $B$ and $T$ dependences are much weaker for $\nu=2/3$ and $5/3$, as well as for the fqH states in our second device—although with a slightly different geometry and measured at higher $T>1.6$ K—whose $p_{CAR}$ are comparable to those for integer states (Figure 4 and 8-14).

### V. GRAPHENE REGION SURROUNDING THE SUPERCONDUCTOR

To investigate the pronounced CAR of particle-like fractions, we simultaneously control top and bottom gate voltage in our second device, which allows for tuning the density in the ~100 nm narrow graphene region surrounding the superconductor without affecting the bulk filling $\nu$ (Figure 4a). This experimental knob affects the filling $\nu'$ around the superconductor, which enables the probing of the profiles of the fqH edges and the extent of charge accumulation at the superconductor interface (contact-induced doping). Varying $\nu'$ for constant $\nu$ changes the existing edge modes along the superconductor, as shown in Figure 4b for the depleted, neutral ($\nu'$ matches $\nu$) and accumulated narrow graphene region. First, we observe that the particle-like fractional states $v$=1/3, 2/5 (Figure 4c) produce a qualitatively different behaviour from the integer (Figure 4e) or hole-conjugate fractional states $v$=3/5, 2/3 (Figure 4d): while CAR in the particle-like fractions is limited to the matched regime, CAR in the integers or hole-conjugate fractions is not suppressed in the accumulation regime. This could be related to the details of edge profile and reconstruction in, for instance, $v$=1/3 and 2/3, where 1/3 is described by a single edge mode whereas 2/3 is described by counterpropagating edge modes of 1 and 1/3 which equilibrate[41,42]. Second, for all fqH states a small decrease of $v'$ compared to $v$ suppresses CAR. These observations are compatible with the scenario that the accumulation region in the graphene channel related to contact-induced doping is very narrow ($\ll$ 100 nm) and thus fully proximitized, which can potentially explain the strong CAR response for particle-like fqH edges. However, we cannot rule out alternative explanations such as a mundane disorder-induced origin for our pronounced CAR in certain fractions.

## VI. TRANSPORT SPECTROSCOPY AND TEMPERATURE DEPENDENCE

Next, we perform spectroscopy in fractional $v$ by varying $V$ and monitoring $p_{CAR}$ and $R_{xx}$ simultaneously at different temperatures. Figure 5a-d show CAR for $v$=1/3 and 2/5, both limited to an energy range below $|eV|$~1 meV and to low $T$ (see also the color plots as insets and Figure 15 which shows $R_{CAR}$ instead of $p_{CAR}$). Increasing $V$ and $T$ bring the injected charges above the excitation gap of the fractional fillings (bulk conduction) or above $\Delta_{ind}$ (Bogoliubov-quasiparticle transport without Andreev reflection), both suppressing CAR. The comparison of fractional $v$ (excitation gaps in the same range as $\Delta_{ind}$) with $v$=2 (largest Landau gap, significantly larger than $\Delta_{ind}$) suggests that for our fractional $v$ an increasing $V$ suppresses CAR primarily due to bulk conduction, while the suppression with $T$ is due to both bulk conduction and Bogoliubov-quasiparticle transport (Figure 5a, c vs. Figure 5a left inset; Figure 5b, d vs. Figure 14).

We proceed with the temperature dependence of $p_{CAR}$ for several integer and fractional $v$ (Figure 5e). We find consistent CAR below ~5 K for all well-developed fillings, demonstrating the robustness of superconductivity in the fqH state. Above this $T$ increasing quasiparticle transport overcomes CAR. The CAR at $v$=1 and 2 persists up to a larger $T$ in our second device measured using a different setup (Figure 14), indicating that quasiparticle transport can be further decreased.

Figure 5f shows the vertical cuts from Figure 5e for several $v$. We find a $p_{CAR}$ saturating at low temperatures for the integer fillings $v$=1, 2 as well as $v$=2/3. Interestingly, a topological $\Delta_{ind}$ is also expected to provide a temperature-independent behaviour for integer $v$, equivalent to the $B$ independence presented in Figure 2 and 3. In stark contrast, the particle-like fillings $v$=1/3 and 2/5 show a clear temperature dependence down to the lowest $T$ (Figure 16), with the $p_{CAR}$ of $v$=1/3 reaching above 6% at $T$=15 mK as shown in Figure 3 ($B$=9 T).

An increasing $p_{CAR}$ with decreasing $T$ could be a feature of the superconducting pairing of fractional charges $e^*$ (Figure 5f right inset and Figure 27). Here, considering $v$=1/3, CAR converts an incoming electron-like $e^*$=1/3 to an outgoing hole-like $-e^*$=-1/3 adding a 2/3 charge to the superconductor. Such $T$ dependence is not expected for pairing of integer charges in fqH edges (Figure 5f left inset), where, for $v$=1/3, three incoming $e^*$ bunch together, which are then converted to three bunched $-e^*$ leaving the superconductor. In this case, CAR vanishes ($R_{CAR}$=0) at zero temperature which is not observed for both our devices at our lowest $T$. Although our experiments do not constitute direct evidence for pairing of fractional charges, which can be complicated by contact-induced doping, our $p_{CAR}$ for fully developed particle-like fqH states being larger than that of integer and hole-conjugate fractional $v$ suggests a different underlying mechanism for pronounced fractional CAR. Direct evidence for fractional charge pairing could be obtained in measurements which are sensitive to the charge of the Andreev-reflected particles.

Our presented experiments show Andreev reflection in the fractional quantum Hall state in the superconductor. Following experiments, including tunneling[28,43], noise[44,45] and supercurrent[46] measurements, will be able to reveal direct evidence for pairing of fractional charges in this hybrid system. Accessing the topological properties will require significant increase of CAR probability by a suppression of edge disorder and vortices.

## ACKNOWLEDGEMENTS

We thank Gil-Ho Lee, Yuval Oreg, Srijit Goswami and Antonio Manesco for helpful discussions. Ö.G. acknowledges support by a Rubicon grant of the Netherlands Organization for Scientific Research (NWO). Y.R. and P.K. acknowledge support from DOE (DE-SC0012260) for measurement and analysis, and NSF (QII-TAQS MPS 1936263) for characterization. S.Y.L. and Y.H.L. acknowledge support from the Institute for Basic Science (IBS-R011-D1). H.S. and A.V. acknowledge support by the Simons Collaboration on Ultra-Quantum Matter, which is a grant from the Simons Foundation (651440, A.V.). J.Z. and A.Y. acknowledge support from DOE (DE-SC0019300) for device fabrication. A.Y. acknowledges funding by the NSF DMR-1708688. P.K. and A.Y. acknowledge experimental collaboration support from Science and Technology Center for Integrated Quantum Materials, NSF Grant No. DMR-1231319. K.W. and T.T. acknowledge support from the Elemental Strategy Initiative conducted by the MEXT, Japan (Grant Number JPMXP0112101001) and JSPS KAKENHI (Grant Numbers 19H05790 and JP20H00354). Nanofabrication was performed at the Center for Nanoscale Systems at Harvard, supported in part by an NSF NNIN award ECS-00335765.

Ö.G., Y.R. and P.K. conceived the idea and designed the project. P.K. supervised the project. Ö.G., Y.R., S.Y.L. and J.Z. fabricated the devices. H.S. and A.V. set up the theoretical model. Y.H.L. and A.Y. consulted on and helped in different stages of the fabrication process and analysis. Ö.G., Y.R. and J.Z. performed the measurements. Ö.G., Y.R., H.S., J.Z., A.Y. and P.K. wrote the paper with inputs from all authors.

## APPENDIX A: EXPERIMENTAL METHODS

### 1. Assembly of the heterostructure

We assembled our five-layer graphite/hexagonal boron nitride (hBN)/single layer graphene/hBN/graphite van der Waals heterostructures with the standard dry transfer technique[47], using a polycarbonate (PC)/polydimethylsiloxane (PDMS) stamp. We exfoliated the flakes via thermal release tape onto (doped) Si substrates covered with 285-nm-thick $SiO_2$—the same substrate on which the devices were fabricated. To increase the size of the flakes we baked the substrates, the tape carrying bulk hBN and graphite still adhered, at 100°C for 1 min on a hot plate before releasing the tape. After exfoliation we annealed the substrates with the flakes at high vacuum (~3·10$^{-7}$ mbar) and 350°C for 20 min to remove the tape residues, ramping the temperature from 20°C over three hours. We did not perform these two heat treatments on single-layer graphene to avoid modifying its intrinsic properties. We determined the cleanness and thickness of the flakes by first optical and then atomic force microscopy. The graphite layers in our heterostructures were determined to be ~1.5 - 3 nm thick (5 to 10 layers) while the hBN layers were ~50 nm (top) and 90 - 100 nm (bottom), all confirmed to be atomically flat. We used a thicker bottom hBN to prevent an electrical connection between the bottom graphite and the overlapping superconductor (see Nanofabrication). After choosing suitable flakes, we assembled the heterostructure in a glovebox ($H_2O$<0.1 ppm, $O_2$<0.1 ppm) to decrease contamination between the layers.

To make the stamp, we prepared a PC solution (8 wt.%) and pipetted it onto a glass slide. We placed another glass slide on top of the solution and left the resulting film to cure in ambient conditions. We then placed a rectangular block (~8×5 mm) of PDMS (Gel-Pak) on a separate glass slide and transferred the PC film (with a larger area than that of the PDMS) on top. To ensure adhesion between the PC film and the glass slide we baked the stamp at 180°C for 5 mins, and then mounted the finished stamp on a dry transfer setup.

Before transferring flakes, we flattened the surface of the PC film by touching the stamp onto a bare substrate held by vacuum on a sample stage at 155°C. After a cooldown period during which the PC film detached from the bare substrate, we replaced the substrate with the one containing the top graphite. The stamp was used to pick up the graphite at ~130°C, followed by a cooldown until the stamp detached. We then used this top graphite to pick up the top hBN via van der Waals force. In this step, the top graphite approached the hBN extremely slowly at 150 - 155°C to minimize the formation of bubbles. We picked up this hBN, the graphene, and the bottom hBN with the same procedure described above. We then dropped the whole stack (graphite/hBN/graphene/hBN) from the stamp onto the bottom graphite at ~170°C. Importantly, dropping the stack at this high temperature allowed us to push out the bubbles formed between the layers during assembly[48]. Finally, we released the PC film at 190°C, ending the transfer process. We completed the assembly by removing polymer residues from the heterostructure, first leaving the chip in chloroform overnight and then annealing the heterostructure at high vacuum (~3·10$^{-7}$ mbar) and 350°C for 30 minutes after a 3-hour temperature ramp.

## 2. Nanofabrication

We defined the superconductor, normal leads, and the shape of the device via electron beam lithography followed by reactive ion etching (RIE). For RIE, we used a $CHF_3$, Ar, and $O_2$ gas mixture, whereas we excluded $CHF_3$ for the selective removal of the top graphite. Both the superconductor and the normal leads contact the graphene from its edge, and were deposited following RIE in one lithography step in which the etch mask also served as the deposition mask. For the superconductor, we first selectively etched the top graphite that would otherwise surround the superconductor, leaving a ~100 nm separation to avoid an electrical connection. In the subsequent lithography step, we deposited the superconductor after vertically etching the heterostructure beyond the graphene layer using an etch-stop. This etch-stop is based on the ex-situ measured conductance of a test area of identical layer composition and leaves most of the bottom hBN unetched which insulates the superconductor from the bottom graphite. Unlike the superconductor, the normal leads do not overlap the top or bottom graphite, and instead connect to the fractional quantum Hall heterostructure through graphene that extends beyond both graphite layers (Figure 6).

The normal leads are Cr/Pd/Au (2/7/150 nm), thermally evaporated on a rotating sample stage with ~15° tilt. We deposited our superconductor in an AJA International UHV hybrid system (base pressure ~10$^{-7}$ Torr). The superconductor deposition started with electron beam evaporation of Ti (10 nm, with rotation and 15° tilt) immediately followed by dc magnetron sputtering of Nb/NbN (5/75 nm, without tilt) at a pressure of 3 mTorr and a power of 200 W using a Nb target. We deposited Nb in an Ar environment. For NbN we used an Ar/$N_2$ (50/6 sccm) gas mixture. This superconductor has a critical temperature $T_c$~12 K at $B$=0 T. An overview of the fabricated devices is given in Table 1.

## 3. Measurement

The parts of the graphene that extend beyond the top and bottom graphite, which connect the normal leads to the fractional quantum Hall heterostructure, were doped by using the substrate as a global gate. The top and bottom graphite layers were used as gates to control the charge carrier density in the fractional quantum Hall heterostructure. Device 1 used top graphite as the gate while the bottom graphite was grounded, whereas the opposite was the case for Device 2 except in Figure 4 where Device 2 used top and bottom graphite simultaneously as gates.

We measured Device 1 in a variable temperature inset (VTI) at $T$≥1.75 K and in a dilution refrigerator with its cold finger at $T$=15 mK, and Device 2 in a different VTI at $T$≥1.6 K. For the measurements in VTIs, we used an RC filter attached to the chip carrier, whereas the dilution refrigerator was equipped with RC, copper powder and pi filters all thermalized to the cold finger. We reproduced our observations after every thermocycle for both our devices. Importantly, however, we were unable to measure crossed Andreev reflection without a filter.

We used the standard ac lock-in technique ($f$<100 Hz) with an excitation current $I_{exc}$=5 or 10 nA for the VTI measurements and $I_{exc}$=1 or 5 nA for the dilution refrigerator measurements. All presented measurements used a single source and the superconductor as the single drain. Because of the finite resistance (wiring and filters) between the superconductor and the breakout box at room temperature, $V_{CAR}$ and $V$ (Figure 1b) were measured relative to the superconductor potential. Our superconductor coupled to the fractional quantum Hall edges branches out to four separate leads (Figure 6). Two branch out immediately after leaving the heterostructure (one used as the drain, the other to monitor the potential), and the remaining two (left floating) after a narrow strip with identical dimensions as the one coupled to the fractional quantum Hall edges. This strip allowed for an independent test of our superconductor. An overview of the measured devices is given in Table 2.

## APPENDIX B: THEORY

The inset of Figure 2b shows the schematic of the vertically shrunk experimental system which satisfies $I_{exc} = \nu \cdot e^2/h \cdot (V - V_{CAR})$. A $p_{CAR} = 1$ implies $R_{CAR}/R_{XY} = -1/2$. In what follows, we first present a low energy theory of graphene in the presence of a strong out-of-plane $B$ [49] (Figure 17) and compute crossed Andreev reflection (CAR) for integer fillings. We conclude with the renormalization flow equations which describe CAR response in the fqH edge as the temperature is varied.

We focus on 0<ν≤6 using an effective continuum model, extending the previous tight binding based numerical studies on ν=0[50] and ν=1[51]. The analytic Bogoliubov-Dirac-Landau equation, which agrees with the previous numerical studies[50,51] allows access to systems with a large magnetic length, relevant to the experimental data at higher fillings with multiple qH edge modes. We put the system on a cylinder and choose the Landau gauge to preserve translation invariance along the compactified direction ($y$-axis). This choice is for convenience in our calculations and does not affect the spectral properties or the response functions.

We encapsulate the slowly varying field operators in an eight-component field and introduce the Pauli matrices $s_\alpha$, $\tau_\alpha$ and $\sigma_\alpha$ which act on spin, valley and sublattice degrees of freedom, respectively. In this notation, the Hamiltonian reads $H_0 = \int d^2 x \, \psi^\dagger(r) h_0 \psi(r)$ where

$$h_0 = v_F(-i\hbar\partial_x - eA_x)\tau_z\sigma_x + v_F(-i\hbar\partial_y - eA_y)\sigma_y + m\sigma_z - \mu + gs_z$$

with the chemical potential $\mu$, the Zeeman term $g$, and a Dirac mass term $m$ (necessary to realize $\nu=1$ in our noninteracting analysis). The energy levels are given by $\varepsilon_n = \sqrt{|n|\varepsilon_0^2 + m^2} \pm g$ where $n$ is integer and $\varepsilon_0 = \sqrt{2e\hbar B}v_F$ is the characteristic energy scale of the Landau level spacing for the Dirac Hamiltonian. We model the superconducting region with the pairing terms[52,53] $\psi_\uparrow[\Delta_1\tau_x + \Delta_2\tau_0]\psi_\downarrow$. Both pairing terms are of spin-singlet s-wave type, with the first term describing an inter-valley pairing, and the second term an intra-valley pairing. The inter-valley pairing (first term) does not result in a gap when interfaced with an s-wave superconductor so whether it is singlet or triplet in the valley basis is irrelevant. The intra-valley pairing (second term), on the other hand, is essential for gap opening and present only for the interface along the armchair edge. Therefore, we limit our analysis to an armchair edge (interface along $y$).

To find the energy spectrum at the interface between qH and a superconducting region, we solve the Bogoliubov-de Gennes (BdG) equation in the plane-wave basis $e^{i2\pi kx/W}$ where $W$ is the length of the cylinder and $k = 0, \pm 1, \pm 2, \cdots$ with a momentum cut-off $|k| < N_x$. Figure 18 shows a typical energy spectrum near the edge for such interface. Here we set $g = 0$ and the bands are doubly degenerate due to spin rotation symmetry[54].

We now consider the counterpropagating edge modes along either side of the superconductor (Figure 2b inset, $y$-axis is along the superconductor). For reference, we show in Figure 19 the armchair edge energy spectra in the absence and presence of the Dirac mass term, where we replace the superconducting region with vacuum. Without any (valley) symmetry breaking, the quantum Hall sequence is $\nu = 2,4,6,\cdots$ Adding the Dirac mass term modifies the sequence into $\nu = 1,2,4,\cdots$ with a spin and valley polarized $\nu = 1$. Turning on superconductivity yields Figure 20, with the top panels for $\nu = 1$ and the bottom panels for $\nu = 2$. For any integer filling zero-energy band crossings are between particle-hole partners with identical spin polarization. Therefore, there is no direct mechanism for s-wave pairing (even for spin-unpolarized fillings such as $\nu = 2$). However, the large spin-orbit coupling in NbN superconductor provides a necessary ingredient for a spin-flip process allowing for a pairing between electrons with the same spin polarization. We account for spin-orbit coupling via[55]

$$h_{so} = \lambda_{so}\tau_z s_z\sigma_z + \lambda_{R,x}\tau_z s_y\sigma_x - \lambda_{R,y} s_x\sigma_y.$$

Because the induced pairing between the counterpropagating edge modes exponentially decays as a function of the superconductor thickness, a gap does not open for a thick superconductor (left panels). This is the experimental scenario in which Andreev edge states govern the transport[16,18,29,38,39]. Reducing the thickness of the superconductor hybridizes the edge modes along both sides of the superconductor and opens a gap in the BdG spectrum (middle panels). The band crossings and the resulting hybridization occur for each mode separately—each edge mode (for instance spin-up and spin-down) experiences a similar induced pairing (gap opening) mechanism. The gap opening term is a spin-polarized pairing of the form $\psi_{Ls}\psi_{Rs}$ which originates from the combination of the spin-singlet pairing of the parent superconductor and spin-orbit coupling. Turning off spin-orbit coupling results in zero-energy band crossings remaining intact (right panels). The degeneracy points (band crossings at finite energies) near $k_y = 0$ are lifted for a thin superconductor regardless of the presence of spin-orbit coupling. This is because those degeneracies open by either spin-singlet intra-edge pairing or simply direct tunneling. Nevertheless, these processes do not play any role for the zero-energy band crossings between two spin-polarized modes since the former process does not open a gap in a spin-polarized channel and the latter is forbidden at finite $k$ due to violating momentum conservation. This implies that $\nu = 2$ edge modes can be treated effectively as two copies of $\nu = 1$. Our numerical analysis confirms that the induced gaps within different edge modes are of the same order and decrease as the superconductor is made thicker (Figure 21).

We now proceed to computing the probability of crossed Andreev reflection, a process which we effectively treat as a quasi-1D scattering problem of a superconducting region of width $W_s$ sandwiched between two quantum Hall regions each with a width $W$ (Figure 2b inset, superconductor along y). We first find the transfer operator of Dirac-BdG equation, which is the solution to the wave equation in the Nambu basis

$$\hbar v_F \frac{\partial \Psi_y(x)}{\partial y} = i\eta_z\sigma_y[E - h_0(x) - h_{sc}\,\theta(x)]\Psi_y(x). \quad (1)$$

Here, $h_0$ is the Hamiltonian intrinsic to the qH region
$h_0(x) = v_F(-i\partial_x)\tau_z\sigma_x + ev_F A_y(x)\sigma_y + \eta_z(m(x)\sigma_z - \mu(x) + g(x)s_z)$,
$h_{sc}$ the Hamiltonian of the superconducting region
$h_{sc} = \Delta_1\eta_y\tau_x s_y + \Delta_2\eta_y s_y + \lambda_{so}\eta_z\tau_z s_z\sigma_z + \lambda_{R,x}\tau_z s_y\sigma_x - \lambda_{R,y} s_x\sigma_y$,
$\eta_\alpha$ the Pauli matrices acting in the particle-hole subspace, $\theta(x)$ a box function nonzero only in the superconductor region, and $A_y(x)$ the gauge field

$$A_y(x) = \begin{cases} B(x - W_s/2) & x > +W_s/2 \\ B(x + W_s/2) & x < -W_s/2 \\ 0 & |x| \leq +W_s/2 \end{cases}$$

chosen such that the magnetic field vanishes inside the superconductor, a common simplification[51]. The Dirac mass term, local chemical potential, and the Zeeman term are

$$m(x) = \begin{cases} m_s & |x| \leq W_s/2 \\ m_n & |x| > W_s/2 \end{cases} \quad \mu(x) = \begin{cases} \mu_s & |x| \leq W_s/2 \\ \mu_n & |x| > W_s/2 \end{cases}$$
$$g(x) = \begin{cases} g_s & |x| \leq W_s/2 \\ g_n & |x| > W_s/2. \end{cases}$$

We neglect the Zeeman term in the superconductor for simplicity as it has no effect on the calculated induced pairing gap. We set $\mu_s = 8\varepsilon_0$ and $m_s = 3\varepsilon_0$ to model a superconductor with quadratic bands ($\sqrt{k^2 + m_s^2} = m_s + k^2/2m_s + \cdots$).

We now solve equation (1) in a plane-wave basis representation where the basis vectors are given by

$$\psi_{ek}^{\pm} = \frac{1}{\sqrt{W_t}} e^{iq_k x} |+_z\rangle_\eta |\pm_y\rangle_\sigma, \quad \psi_{hk}^{\pm} = \frac{1}{\sqrt{W_t}} e^{iq_k x} |-_z\rangle_\eta |\mp_y\rangle_\sigma,$$

$$q_k = \frac{2\pi k}{W_t},$$

where $W_t = 2W + W_s$ is the total width of the system, $|\pm_y\rangle_\sigma$ denote eigenstates of $\sigma_y$, that is, $\sigma_y |\pm_y\rangle_\sigma$, and $|\pm_z\rangle_\eta$ denote particle (hole) states, that is, $\eta_z |\pm_z\rangle_\eta = \pm |\pm_z\rangle_\eta$ (recall $\eta_z \equiv 2\psi^\dagger \psi - 1$). We put the subscript $e$ and $h$ to emphasize the distinction between the particle and hole states. We omitted the explicit writing of spin/valley basis vectors in the states above for brevity since the kinetic term in $y$ direction is diagonal in these subspaces. The final quantities involve summations over both spin and valley degrees of freedom. The reflection matrix $r$ is then determined by matching the amplitudes of incoming, reflected, and transmitted modes at $y = 0$,

$$\sum_k [\delta_{kk'} \psi_{ek}^{+}(x) + r_{kk'}^h \psi_{hk}^{-}(x) + r_{kk'}^e \psi_{ek}^{-}(x)] = \Psi_0(x)$$

where $\Psi_0(x)$ only contains forward propagating and decaying evanescent wave solutions to equation (1).

To get the scattering of an incoming electron from the upper qH plane $x > W_s/2$ to a hole in the lower half plane $x < -W_s/2$, we compute the *collective* probability of crossed Andreev reflection $P_h = Tr(r_h^\dagger X r_h X^*)$ (note $0 \le P_h \le \nu$), where $X$ matrix is the unitary matrix transforming $q_k$-modes to real-space modes $X_{kk'} = \int_0^W e^{i2\pi(k-k')(x+W_s/2)/W_t} dx$. Left panel of Figure 22 shows the collective crossed Andreev reflection probability as a function of the chemical potential in the qH region. The plateaus correspond to different integer qH states in the normal region and the quantization is because of gap opening in the spectrum which effectively gives rise to $\nu$ copies of topological superconductors at a filling $\nu$. We also plotted the bias dependence at two values of chemical potential which represent filling $\nu = 1$ and 2. It is interesting to note that there is a second plateau for $\nu = 2$ at finite bias which corresponds to gap opening between opposite spin species (similar to ordinary Andreev reflection).

### 1. Effective edge theory

We now proceed with effective theories at $\nu$ = 1 and 2 edges for a more quantitative understanding. For simplicity, we start with $\nu$ = 1 and derive a minimal model which reproduces the BdG spectrum in Figure 20. To derive an ansatz for nonlinear conductance $G = I_{exc}/V$, we first reduce the problem to a 1D scattering between two sets of edge modes, as shown in inset of Figure 2b. We begin with the effective Hamiltonian

$$H_0 = v_F \int dy \, [\psi_R^\dagger(y)(-i\partial_y - k_0)\psi_R(y) + \psi_L^\dagger(y)(i\partial_y - k_0)\psi_L(y)],$$

where we introduce two Fermi fields $\psi_L$ and $\psi_R$ to describe the edge modes near the upper and lower voltage probes. $R$ and $L$ refer to the right/left movers. We take into account the proximity effect of superconductor via the following terms

$$\Delta \psi_R \psi_L + \sum_{\sigma=L,R} \tilde{\Delta} \psi_\sigma (-i\partial_y) \psi_\sigma + \text{H.c.}$$

where $\Delta$ and $\tilde{\Delta}$ are two phenomenological constants to model the inter-edge and intra-edge pairing, respectively. We neglect direct tunneling between right and left movers—modes with opposite $k$—since it violates momentum conservation along the edges.

It is more convenient to rewrite the full Hamiltonian in the Nambu basis

$\Psi^\dagger = (\psi_R^\dagger, \psi_L^\dagger, \psi_R, \psi_L)$. The problem is then mapped to a single-particle BdG equation $\mathcal{H}|\Psi\rangle = E|\Psi\rangle$,

$$\mathcal{H} = v_F(-i\partial_y)\sigma_z - v_F k_0 \tau_z + \Delta \tau_y \sigma_y + \tilde{\Delta}(-i\partial_y)\tau_x$$

where $\tau_\alpha$ and $\sigma_\alpha$ are Pauli matrices acting on particle-hole and left/right mover degrees of freedom, respectively. This Hamiltonian is particle-hole symmetric, that is, it satisfies the identity $U\mathcal{H}^T U^\dagger = -\mathcal{H}$ where $U = \tau^x$, and belongs to topological phases of the symmetry class D in the noninteracting classification[56–58]. The Bogoliubov spectrum associated with this Hamiltonian is given by

$$\epsilon_q = \pm \sqrt{\left(v_F q \pm \sqrt{\tilde{\Delta}^2 q^2 + v_F^2 k_0^2}\right)^2 + \Delta^2},$$

which is shown in Figure 23. The intra-edge term modifies the velocity of the modes $v_F \to v_F \pm \tilde{\Delta}$. However, crossed Andreev reflection (discussed below) does not depend on this effect.

For simplicity, we assume that the superconductor is infinitely long. We consider the interface to be at $y = 0$, so that

$$\Delta(y) = \begin{cases} \Delta & y > 0 \\ 0 & y < 0. \end{cases}$$

We consider the following ansatz for the wave function in the two regions:

$$\Psi(y) = A_1 \Psi_{ke}^{(R)}(y) + A_2 \Psi_{kh}^{(R)}(y) + B_1 \Psi_{ke}^{(L)}(y) + B_2 \Psi_{kh}^{(L)}(y) \qquad y < 0$$

$$\Psi(y) = C_1 \Psi_{q_+}^{(R)}(y) + C_2 \Psi_{q_-}^{(R)}(y) + D_1 \Psi_{q_+}^{(L)}(y) + D_2 \Psi_{q_-}^{(L)}(y) \qquad y > 0,$$

where the superscript $(R)$ or $(L)$ refer to the right or left movers, respectively, and $h$ and $e$ denote the electron and hole degrees of freedom.

To find the scattering matrix elements, we find the transfer matrix by demanding the wave function to be continuous at the interface $y = 0$:

$$\begin{pmatrix} C \\ D \end{pmatrix} = \begin{pmatrix} T_{RR} & T_{RL} \\ T_{LR} & T_{LL} \end{pmatrix} \begin{pmatrix} A \\ B \end{pmatrix}, \qquad (2)$$

where $T_{ij}$ are $2 \times 2$ matrices in particle-hole basis and the subscripts denote the transfer of right/left mover to left/right mover from $y < 0$ region to $y > 0$ region. We also multiply it by another transfer matrix to account for various processes such as electrons bypassing the superconductor (shown by the arrow in Figure 2b inset) or possible chemical potential or velocity mismatch[59] between incoming modes and hybridized modes,

$$T^{(z)} = \begin{pmatrix} 1-iZ & 0 & -iZ & 0 \\ 0 & 1+iZ & 0 & -iZ \\ iZ & 0 & 1+iZ & 0 \\ 0 & iZ & 0 & 1-iZ \end{pmatrix},$$

which is parametrized by $Z$ as follows: $Z = 0$ and $Z \to \infty$ correspond to no tunneling (perfect junction with the proximitized modes) and perfect tunneling (fully detached from the proximitized modes), respectively.

From the above relation, we can read off the scattering matrix for an incoming wave such as

$$\Psi_{in}^e(y) = \frac{1}{\sqrt{2\pi v_F}} \begin{pmatrix} |+\rangle \\ 0 \end{pmatrix} e^{ik_e y},$$

which corresponds to

$$A = \begin{pmatrix} 1 \\ 0 \end{pmatrix}, \quad B = \begin{pmatrix} r_{ee} \\ r_{he} \end{pmatrix}, \quad C = \begin{pmatrix} t_{ee} \\ t_{he} \end{pmatrix}, \quad D = \begin{pmatrix} 0 \\ 0 \end{pmatrix}.$$

The reflection coefficients can then be found by eliminating $C$ from equation (2),

$$B = \begin{pmatrix} r_{ee} \\ r_{he} \end{pmatrix} = -T_{LL}^{-1} T_{LR} \begin{pmatrix} 1 \\ 0 \end{pmatrix}.$$

In the clean limit, we exactly recover the Blonder-Tinkham-Klapwijk (BTK) formulas[35]. In the zero-bias limit $E = 0$ and for arbitrary $Z$, we have

$$r_{ee} = -\frac{2Z(Z+i)}{2Z^2 + 1}, \quad r_{he} = \frac{i}{2Z^2 + 1},$$

which are independent of Δ.

To model a realistic system, we consider random pairing terms
$$V_{dis} = \Delta(y)\tau^y\sigma^y$$
which varies uniformly in the range $[-\Delta_m, \Delta_m]$ over distances comparable to the minimum coherence length $\xi_0 \sim v_F/\Delta_m$. Although this is a heuristic ansatz for an actual experimental setup, we motivate our model by noting that the leading order effect of having finite chemical potential disorder within the superconductor or near the junction and the presence of magnetic flux vortices within the superconductor cause variations in the induced pairing in the edge modes.

## 2. CAR resistance

Using the reflection coefficients, we can then find the differential conductance,
$$G(V) = \frac{dI_{exc}}{dV} = \frac{e^2}{h}(1 - |r_{ee}|^2 + |r_{he}|^2).$$
In the large bias regime, it is simplified into
$$G(eV \gg \Delta, \Gamma) \approx \frac{e^2}{h}\frac{1}{1+Z^2},$$
similar to the regular BTK case. In the linear regime, we get
$$G(0) = \frac{2e^2}{h}\frac{1}{(1+2Z^2)^2},$$
which is identical to the conventional BTK, and independent of Δ. Using the fact that in this regime $I_{exc} = G(0)V$ together with $G_{XY} = e^2/h$ we arrive at
$$R_{CAR} = \frac{h}{2e^2}((1+2Z^2)^2 - 2),$$
which implies that $Z < Z_c = (1/\sqrt{2} - 1/2)^{1/2} \approx 0.45$ is necessary to get a negative $R_{CAR}$.

Now we study the $\nu = 2$ integer quantum Hall state. The edge theory is described by two chiral modes associated with the two spin degrees of freedom,
$$H_{edge} = v_F \sum_{s=\uparrow,\downarrow} \int dx\, \psi_s^\dagger(x)(-i\partial_x - k_{Fs})\psi_s(x),$$
where $v_F$ is the Fermi velocity, and $k_{F\uparrow}$ and $k_{F\downarrow}$ denote the Fermi wave vector for spin up and down.

Again, we first reduce the problem to a 1D scattering between two types of edge modes, as shown in Figure 2b (inset). Putting the two edge modes together, we write an effective Hamiltonian,
$$H_0 = v_F \sum_{s=\uparrow,\downarrow} \int dy\, [\psi_{Rs}^\dagger(y)(-i\partial_y - k_{Fs})\psi_{Rs}(y) + \psi_{Ls}^\dagger(y)(i\partial_y - k_{Fs})\psi_{Ls}(y)],$$
where we introduce two Fermi fields $\psi_{Ls}$ and $\psi_{Rs}$ to describe the upper and lower edge modes. Here, $s = \uparrow, \downarrow$ denotes the electron spin. We consider the proximity effect of superconductor via the following terms
$$\sum_{s=\uparrow,\downarrow}(\Delta_1 s\, \psi_{Rs}\psi_{L\bar{s}} + \Delta_2 \psi_{Rs}\psi_{Ls}) + \tilde{\Delta}_1\sum_{\sigma=L,R}\psi_{\sigma\uparrow}\psi_{\sigma\downarrow} + \text{H.c.}$$
Here, $\Delta_1$ and $\Delta_2$ terms describe spin singlet and triplet inter-edge induced pairing while $\tilde{\Delta}_1$ describes a spin singlet intra-edge pairing, respectively. We should note that the pairing potentials $\Delta_1$ and $\tilde{\Delta}_1$ are between spin up and down electrons while the $\Delta_2$ pairing preserves the spin polarization. Therefore, we need spin-orbit coupling to generate the latter term as opposed to the former terms which are induced by the original s-wave pairing in the superconductor.

It is more convenient to rewrite the full Hamiltonian in the Nambu basis $(\Psi_\uparrow^\dagger, \Psi_\downarrow^\dagger)$ where $\Psi_s^\dagger = (\psi_{Rs}^\dagger, \psi_{Ls}^\dagger, \psi_{Rs}, \psi_{Ls})$. The problem can then be mapped to a single-particle BdG equation $\mathcal{H}|\Psi\rangle = E|\Psi\rangle$,

$$\mathcal{H} = v_F(-i\partial_y)\sigma_z + \sum_s k_{Fs}\tau_z|s\rangle\langle s| + \Delta_1\tau_y\sigma_x s_y + \tilde{\Delta}_1\tau_y s_y + \Delta_2\tau_y\sigma_y,$$

where $\tau_\alpha, \sigma_\alpha, s_\alpha$ are Pauli matrices acting on particle-hole, left/right mover, and spin up/down degrees of freedom, respectively. The Bogoliubov spectrum of this system for two cases of parameters is plotted in Figure 24. Note that $\tilde{\Delta}_1$ does not open a gap in the spectrum and only shifts the zero-energy band crossings horizontally, and the gap opening between $\psi_{L\uparrow}$ and $\psi_{R\downarrow}^\dagger$ at finite energy is mediated by the s-wave induced pairing $\Delta_1$. It is within this energy gap that we obtain $P_h = 1$ in Figure 24 (right panel).

To study crossed Andreev reflection, we consider a disordered pairing potential
$$V_{dis} = \Delta_1(y)\tau^y\sigma^x s^y + \Delta_2(y)\tau^y\sigma^y$$
which we vary over the minimum coherence length $\xi_0 \sim v_F/\Delta_m$. Here, both $\Delta_1$ and $\Delta_2$ are random variables drawn from a uniform distribution $[-\Delta_m, \Delta_m]$. The resulting CAR is shown in Figure 25. It is evident that in both clean and disordered case the zero-bias $R_{CAR}$ is inversely proportional to the filling $\nu$, consistent with the experiment. Notably, the experimental observation of CAR vanishing at finite energies can only be recovered in our calculation by introducing disorder which breaks translation symmetry and effectively produces direct tunneling terms. For the clean case, CAR remains consistently finite even at high energies. This implies that the pairing gap and disorder together determine the energy range of CAR. Finally, Figure 26 compares the calculated CAR response with the experiment, qualitatively reproducing the observed difference between ν=2 and 6. The larger energy range of CAR in ν=2 is due to its counterpropagating edge modes being closer to the superconductor than those of ν=6, which results in a stronger hybridization (larger $\Delta_{ind}$).

## 3. FqH

Now we turn to fqH, which requires different approaches from the integer qH CAR discussed above. For simplicity, we shall consider only the $\nu = 1/m$ Laughlin states (with odd $m$), where the edge theory is single component. The system is described by two chiral edge modes near top and bottom of the sample:
$$\mathcal{L}_{edge} = \frac{1}{4\pi\nu}\int dy\, [\partial_t\phi_L \partial_y\phi_L - V\partial_y\phi_L\partial_y\phi_L]$$
$$+ \frac{1}{4\pi\nu}\int dy\, [-\partial_t\phi_R\partial_y\phi_R - V\partial_y\phi_R\partial_y\phi_R]$$
where
$$[\phi_{L,R}(y_1), \phi_{L,R}(y_2)] = \mp i\frac{\pi}{m}\text{sgn}(y_1 - y_2),$$
$$[\phi_L(y_1), \phi_R(y_2)] = i\frac{\pi}{m}.$$
$\phi_{L,R}$ are chiral bosons[60,61]. Using this formulation, the electric charge density associated with $\phi_\alpha$ is given by
$$\rho_\alpha = \frac{1}{2\pi}\partial_y\phi_\alpha,$$
and the $I$th electron operator is described by the vertex operator
$$\Psi_{e,L/R} = e^{i\nu^{-1}\phi_{L/R}}.$$
In the right half of the system (recall Figure 2b inset) the two edge modes are decoupled. In the left half, the two edge modes are coupled via the following induced pairing term
$$\mathcal{L}_{pairing} = \Delta \int_0^\infty dx \cos\nu^{-1}(\phi_L + \phi_R).$$
Note that we drop the Klein factor since it commutes with other terms in the Hamiltonian. The bare scaling dimension of this term is $1 - \nu^{-1}$. However, we work in the strong coupling limit $\Delta \to \infty$. The ground state of the $x > 0$ region is obtained by pinning the field $\nu^{-1}(\phi_L + \phi_R) =$

$2\pi n$ where $n = 0, 1, 2, \cdots, \nu^{-1} - 1$. This in turn implies not only $\langle e^{i\nu^{-1}(\phi_L + \phi_R)} \rangle \neq 0$ but also $\langle e^{i(\phi_L + \phi_R)} \rangle \neq 0$, that is, we get a condensate of quasi-particle and quasi-hole pairs. The scattering of a (quasi)particle impinging on the $x > 0$ region from the left region can be addressed by the following two processes:

1. Coherent conversion of a right-moving quasi-electron to a left-moving quasi-hole, which is described by the following term

$$\Gamma_{qp} \int dy\, \delta(y - 0^-)\, e^{i(\phi_L + \phi_R)} + \text{H.c.}$$

2. Coherent conversion of a right-moving electron to a left-moving hole, which is described by the following term

$$\Gamma_{eh} \int dy\, \delta(y - 0^-)\, e^{i\nu^{-1}(\phi_L + \phi_R)} + \text{H.c.}$$

The latter process requires electron bunching before going through the superconductor. The scaling dimensions of the two processes are given by $\nu$ and $\nu^{-1}$, respectively. Hence, the tunneling amplitudes obey the renormalization flow equations

$$\frac{d\Gamma_{qp}}{dl} = (1 - \nu),$$
$$\frac{d\Gamma_{eh}}{dl} = (1 - \nu^{-1}),$$

which implies that in the infra-red limit (low energy), for $\nu < 1$, $\Gamma_{qp}$ is relevant while $\Gamma_{eh}$ is irrelevant. Now, we compare experimental consequences of both processes (Figure 27):

1. $\Gamma_{qp}$: Since this term is relevant, it drives the system to a strong coupling limit fixed point $\Gamma_{qp} \to \infty$ in the low-energy limit. So, we expect that at low temperatures there is a constant CAR response proportional to $|\Gamma_{qp}|^2$ and it gradually decreases as $T^{2\nu-2}$ when the temperature is increased or as $V^{2\nu-2}$ when a bias voltage is applied.

2. $\Gamma_{eh}$: Since this term is irrelevant, at zero temperature the system in a weak coupling limit $\Gamma_{eh} \to 0$ and there is no CAR response. We expect that as we increase the temperature or apply a bias voltage, the CAR response increases as $T^{2\nu^{-1}-2}$ or $V^{2\nu^{-1}-2}$, respectively.

Given that the experimental data as a function of temperature shows an initial plateau and gradual increase as $T$ is decreased, we conclude that $\Gamma_{qp}$ is the dominant term governing the CAR response. Further quantitative discussion on fqH CAR requires many-body simulation of the fqH states.

____________________

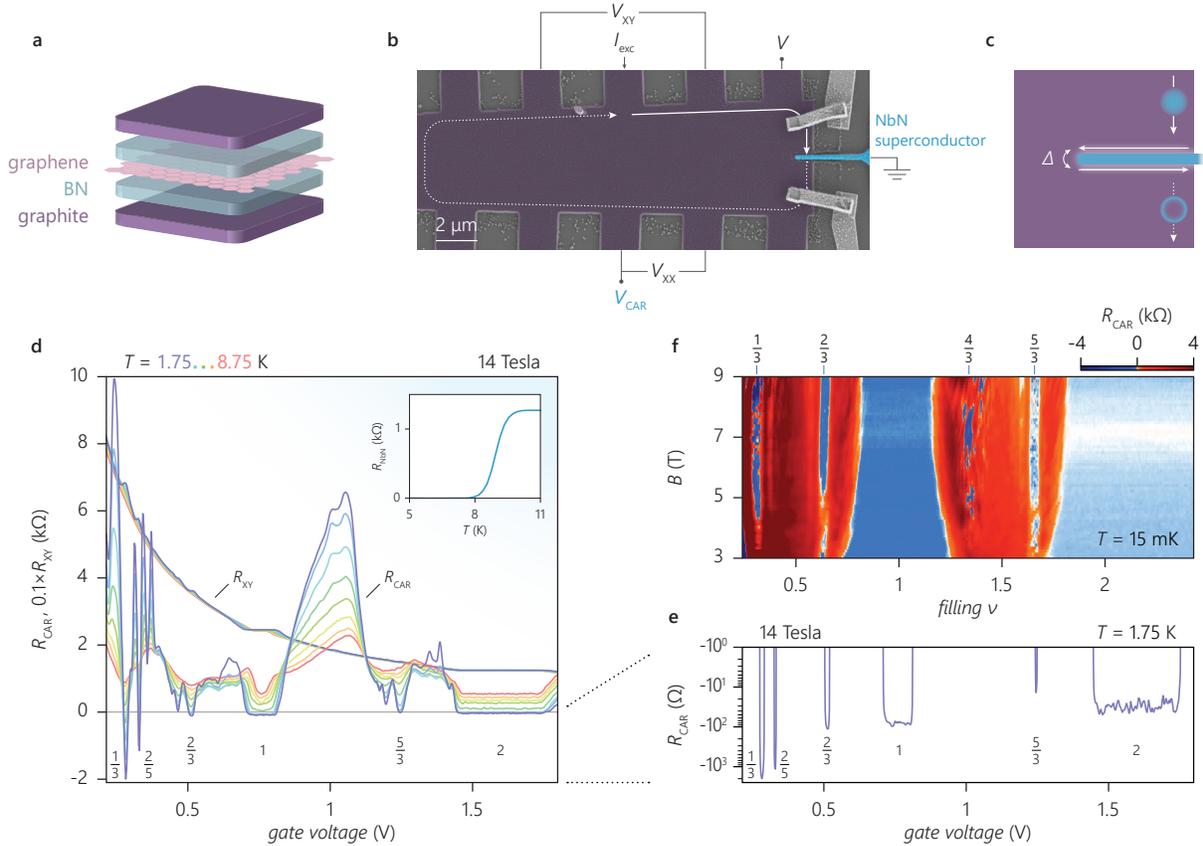

FIG. 1. **The device and crossed Andreev reflection in fqH. a,** Schematic of the heterostructure. Graphene is encapsulated with boron nitride dielectric and graphite. **b,** Typical device including a NbN superconductor <100 nm in width and ~1 μm in length (blue). The extended arms connect to normal leads (not shown) that are used to bias a current $I_{exc}$, measure the voltages $V_{XX}$ and $V_{XY}$, the potential of the edge mode propagating towards the superconductor $V$, and finally that of the edge mode propagating away $V_{CAR}$. The superconductor is grounded, remaining leads are floating. The solid and the dashed arrows depict respectively the chiral electron and hole conduction in an out-of-plane magnetic field $B$. The metal electrodes on top graphite (gate) are bridges that connect the top gate to leads avoiding the edge of the heterostructure. **c,** Illustration of the theory model. A narrow superconductor induces a pairing gap $\Delta$ between the counterpropagating fractional quantum Hall edges along both sides. $\Delta$ converts an incoming electron to an outgoing hole by crossed Andreev reflection (CAR). **d,** $R_{CAR}=V_{CAR}/I_{exc}$ and $R_{XY}=V_{XY}/I_{exc}$ as a function of gate voltage measured at $B=14$ T for different temperatures $T$. An $R_{CAR}<0$ at fractional quantum Hall plateaus indicates hole conductance (CAR). **Inset of d** shows the resistance of a narrow NbN for varying $T$ which superconducts below 8 K at 14 T. **e,** $R_{CAR}$ at 1.75 K from **d**. **f,** $R_{CAR}$ as a function of filling $\nu$ and $B$ measured at 15 mK. CAR ($R_{CAR}<0$) is observed at 3 T for $\nu=2/3$.

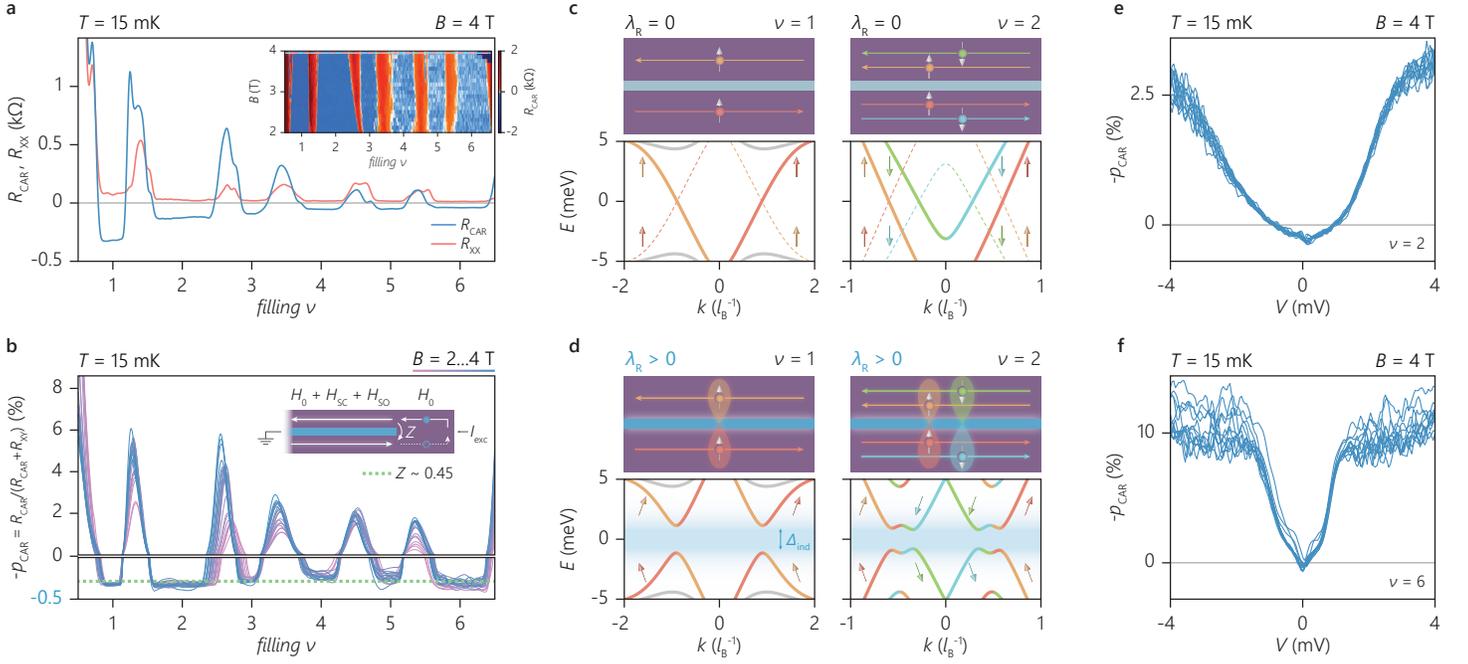

FIG. 2. **Spin-orbit coupling. a,** $R_{CAR}$ and $R_{XX}$ as a function of filling $\nu$. All integer $\nu$ show crossed Andreev reflection (CAR, $R_{CAR}<0$). Small $R_{XX}$ indicates negligible bulk conductance. **Inset of a** shows $R_{CAR}$ measured at $B=2...4$ T. **b,** CAR probability $p_{CAR}$ of the **inset of a**. All integer $\nu$, including the spin-polarized $\nu=1$, have a comparable $p_{CAR}$, an evidence for the pairing mechanism being the same for all integer $\nu$, enabled by spin-orbit coupling. The dashed line is the $p_{CAR}$ when incoming charges can tunnel to the outgoing edge mode without Andreev reflection, represented by $Z$, shown in **inset**. $Z\sim0.45$ matches the measurement. $H_0$ is the Hamiltonian of the edges, $H_{SC}$ the pairing, and $H_{SO}$ the spin-orbit Hamiltonian. **c,** Illustration of the edges separated by a superconductor without spin-orbit coupling ($\lambda_R=0$) for $\nu=1, 2$, and their calculated Bogoliubov-de Gennes spectrum. No pairing gap is present ($\Delta_{ind}=0$). Momentum $k$ in units of $l_B^{-1}$ with $l_B$ the magnetic length. Solid lines are the electron-like excitations, dashed lines the hole-like. Color code indicates the spin and the direction of propagation. **d,** Inclusion of spin-orbit coupling tilts the spins and enables a pairing gap $\Delta_{ind}$. The only possible pairing is p-wave for any integer $\nu$ irrespective of whether it is spin-polarized or not. The inner edge modes pair more strongly due to their proximity to the superconductor. However, for low energy and temperature, $p_{CAR}$ is expected to be independent of the size of $\Delta_{ind}$. **e,** $p_{CAR}$ at $\nu=2$ as a function of the incoming edge mode potential $V$ (excitation) for different gate voltages spanning the entire qH plateau region. Crossed Andreev reflection (CAR) is limited to below $|eV|\sim1$ meV. **f,** Same as **e** but at $\nu=6$. CAR is observed for $|eV|<0.2$ meV.

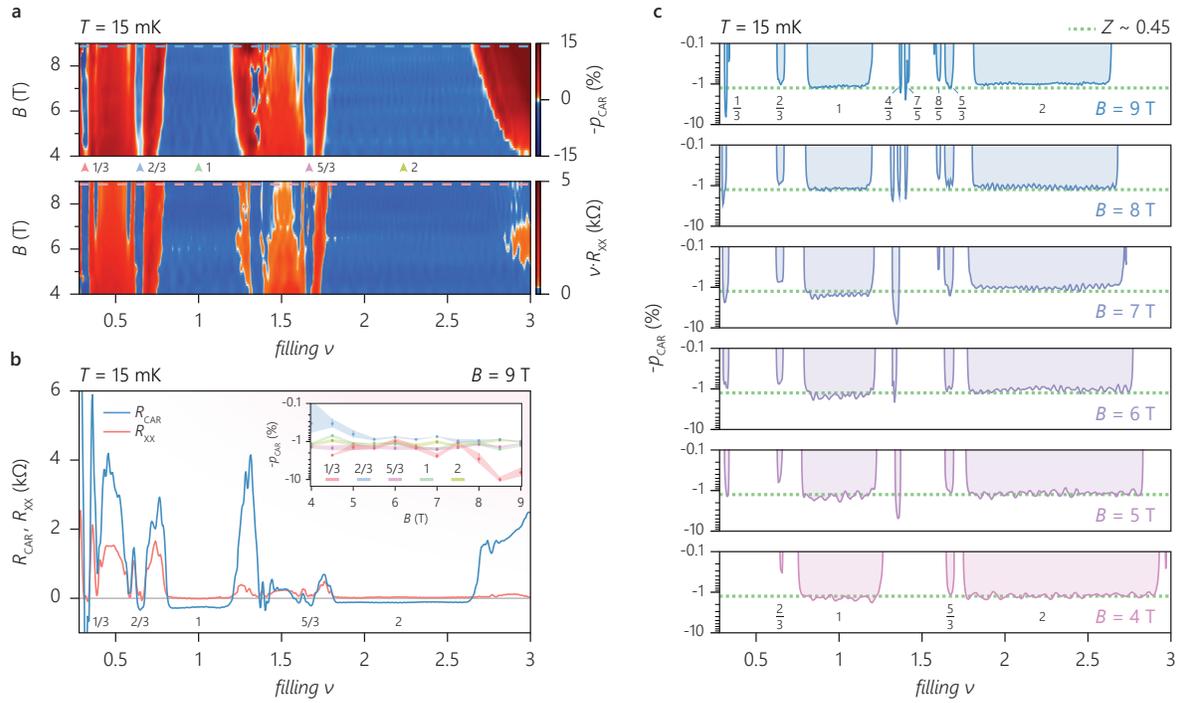

FIG. 3. Magnetic field dependence. **a,** $p_{CAR}$ and $R_{XX}$ (normalized for different $\nu$) as a function of filling and magnetic field $B$. **b,** $R_{CAR}$ and $R_{XX}$ at $B=9$ T from **a**. Small $R_{XX}$ indicates negligible bulk conductance. **Inset of b** shows $p_{CAR}$ from **a** as function of $B$ for several $\nu$. $p_{CAR}$ of $\nu=1/3$ shows a strong $B$ dependence reaching 10% at $B=8.5$ T. $p_{CAR}\sim1\%$ for the rest of the fillings shown. The shades represent the uncertainty in the measured values. **c,** $p_{CAR}$ from **a** for $B=4...9$ T. $p_{CAR}$ at $\nu=1/3$ and $4/3$ changes with $B$, reaching respectively >6% ($B=9$ T) and >8% ($B=7$ T). The rest of the fillings are insensitive to $B$ once they are well developed. $Z\sim0.45$ matches the measurement for fillings which do not exhibit a $B$ dependence.

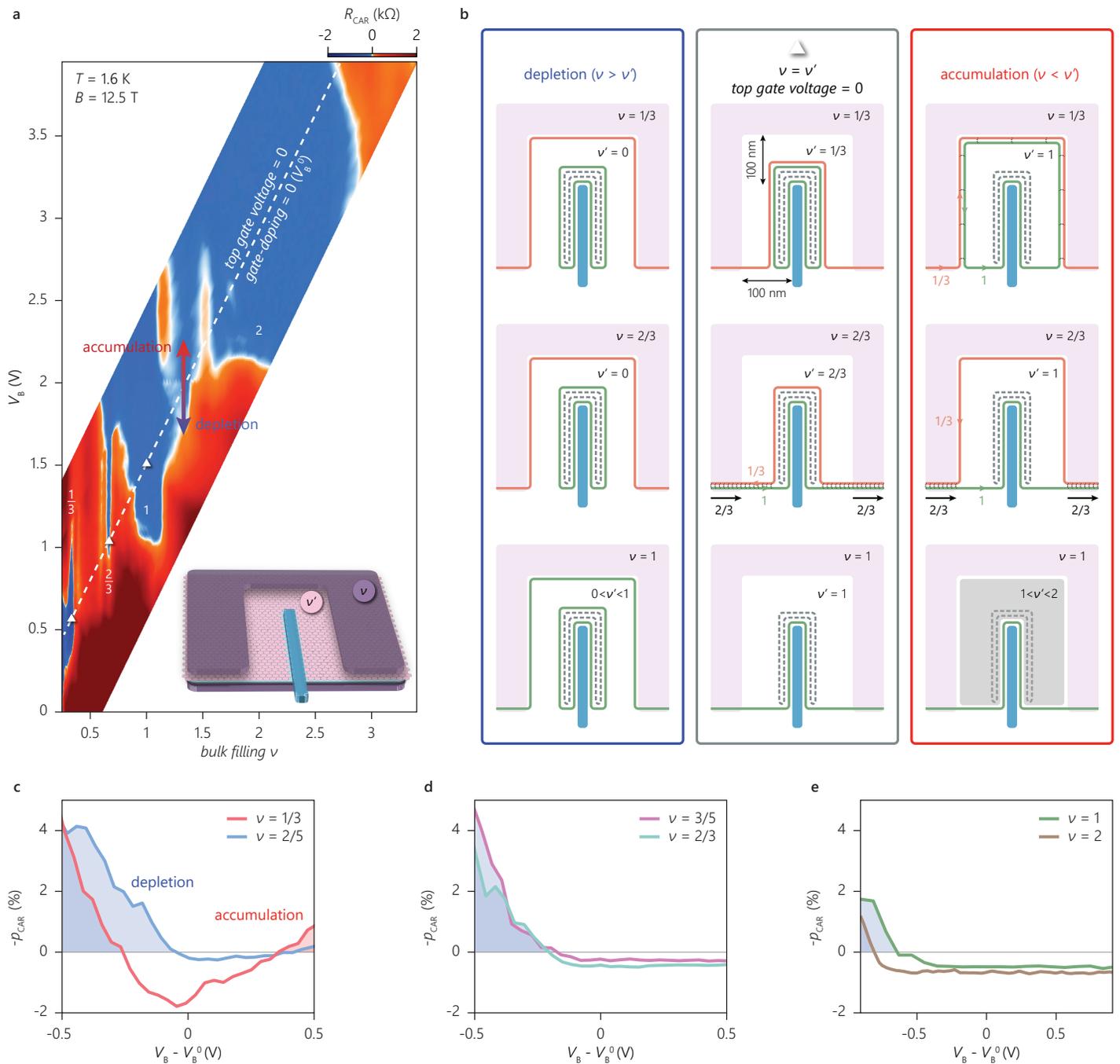

FIG. 4. Graphene region surrounding the superconductor. a, Inset illustrates the region around the superconductor in our devices. The ~100 nm narrow graphene region surrounding the superconductor is not covered by top graphite (gate). This enables tuning the density (filling $v'$) around the superconductor without affecting the bulk filling $v$. Here, $v$ is set simultaneously by top and bottom gate: $v \propto (V_T + c \cdot V_B)$ where $V_T$ is the top gate voltage, $V_B$ the bottom, and $c=0.51$ the ratio of the capacitive couplings of top and bottom gate to the bulk graphene. In contrast, graphene surrounding the superconductor is capacitively coupled primarily to the bottom gate ($v' \propto V_B$). Color plot shows CAR for several fillings $v$ as a function of $V_B$. The symbols mark the edge mode configurations illustrated in b corresponding to the matched ($v=v'$) regime for which $V_T = 0$. Increasing $V_B$ results in the accumulation ($v<v'$) regime, whereas decreasing $V_B$ results in the depletion ($v>v'$) regime. The dashed line marks $V_B^0$, the value of $V_B$ for which $V_T = 0$. b, Illustrations of the edge mode configuration of the particle-like $v=1/3$, the hole-conjugate $v=2/3$ and the integer $v=1$ for varying $v'$. The dashed and the solid green line surrounding the superconductor illustrate integer edge modes potentially induced by an increased charge carrier density at the superconductor interface (contact-induced doping). The thin dashed arcs between the two edge modes 1/3 (orange line) and 1 (green line) depict their equilibration with each other. CAR is absent in the depletion regime, present in the matched regime. Our observed suppression of CAR in $v=1/3$ for $v'=1$ suggests an incomplete equilibration between the edge modes 1/3 and 1 along the electrostatically defined edge (accumulation, top panel). In contrast, CAR in $v=2/3$ is present for $v'=1$, which is consistent with a complete equilibration between the two edge modes along the physical sample edge (accumulation, middle panel). Shaded area in the bottom right accumulation panel depicts a compressible region. The presence of CAR for this filling configuration ($v=1<v'$) suggests negligible tunneling between the edge mode and the compressible region. c, CAR in the particle-like fractional states is limited to small $|V_B - V_B^0|$. d, e, CAR in the integer and the hole-conjugate fractional states are not suppressed for large $V_B - V_B^0$.

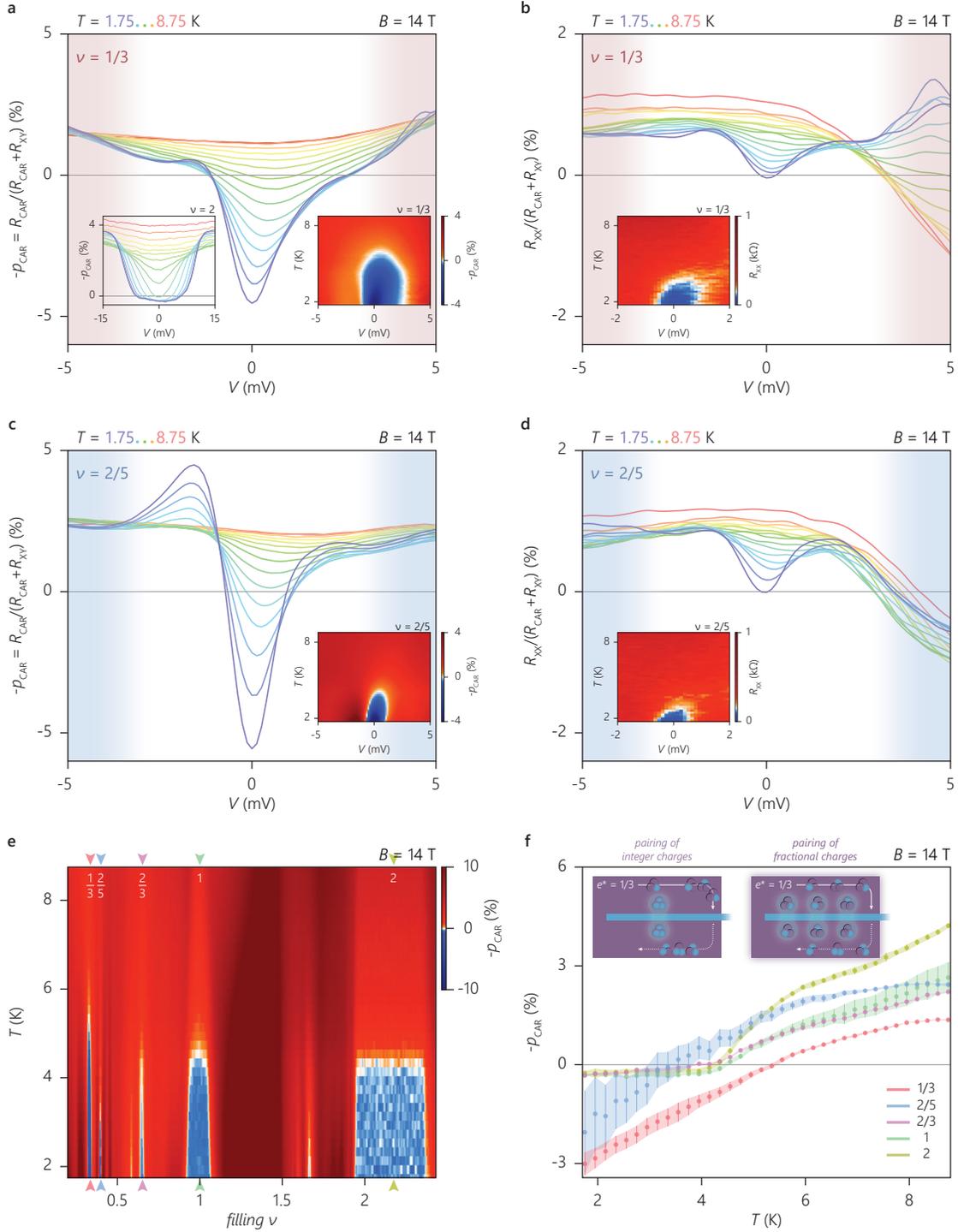

FIG. 5. **Transport spectroscopy and temperature dependence. a,** $p_{CAR}$ at $\nu=1/3$ as a function of the incoming edge mode potential V (excitation) for different temperatures T. Crossed Andreev reflection (CAR) is limited to below $|eV|\sim 1$ meV and T<6 K. Increasing excitation and T suppress CAR. **Left inset of a** shows $p_{CAR}$ at $\nu=2$ with a V and T dependence similar to that at $\nu=1/3$ apart from a larger V range and a smaller $p_{CAR}$. **Right inset of a** shows the $p_{CAR}$ corresponding to **a**. **b,** $R_{XX}$ corresponding to **a**, divided by the incoming edge mode potential $R_{CAR}+R_{XY}$, allowing a direct comparison to $p_{CAR}$. Increasing excitation and T result in bulk conductance which suppresses CAR. **Inset of b** shows $R_{XX}$ corresponding to the **right inset of a**. **c, d,** Same as **a** and **b** but for $\nu=2/5$ which shows a V and T dependence similar to that for $\nu=1/3$. **e,** $p_{CAR}$ as a function of $\nu$ for varying T. $R_{CAR}<0$ below $\sim 5$ K for all well-developed fillings (highlighted with arrows). **f,** Vertical cuts from **e**. At $\nu=2/3$, 1, and 2 $p_{CAR}$ saturates below $\sim 4$ K with decreasing T, whereas at $\nu=1/3$ and 2/5 continues to increase in amplitude without saturating. The shades are the uncertainty in the measurement while sweeping the gate voltage. The uncertainty is larger for fillings represented by fewer data points. **Inset of f** illustrates two different mechanisms of charge transport to the superconductor. Left schematic illustrates bunching of fractional charges of $e^*=1/3$ to form integer charges of e that pair. This mechanism converts three incoming $e^*$ to an outgoing -e, an integer-charged hole, adding 2e to the superconductor. Right schematic illustrates the pairing of fractional charges, a mechanism that converts $e^*$ to $-e^*$, adding 2e/3 to a fractional topological superconductor. For pairing of integer charges, $p_{CAR}$ vanishes at T=0. In contrast, $p_{CAR}$ monotonically increases in amplitude with decreasing T when fractional charges pair.

Device 1

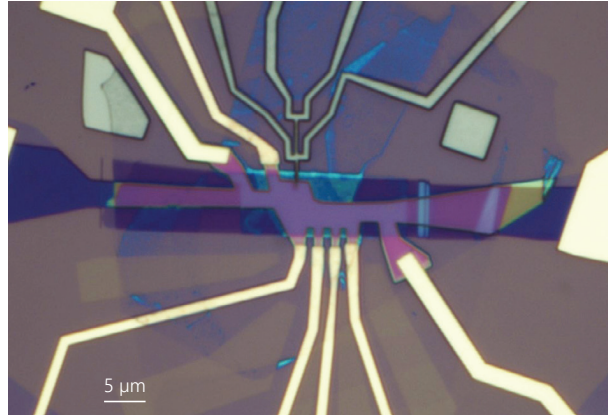

Device 2

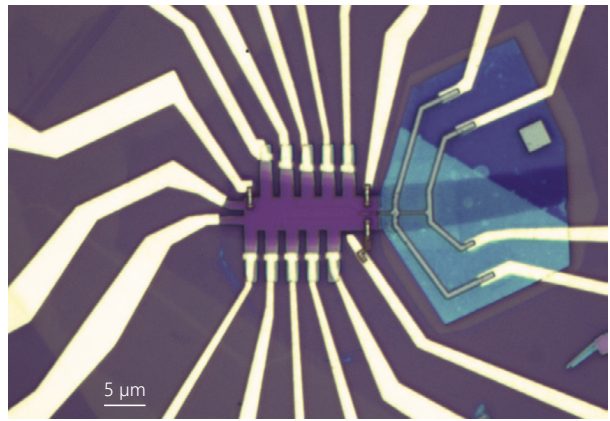

**FIG. 6. The presented devices.** Data presented in Figure 1-3, 5, 15, 16 are taken from Device 1; Figure 4 and 8-14 are taken on Device 2. The superconductor in Device 1 is ~1 μm long resulting in a ~2 μm long graphene–superconductor interface. The superconductor in Device 2 is ~2 μm long which results in a ~4 μm long interface. No systematic dependence of $p_{CAR}$ on the interface length has been observed. The width of the superconductor is <100 nm in both devices.

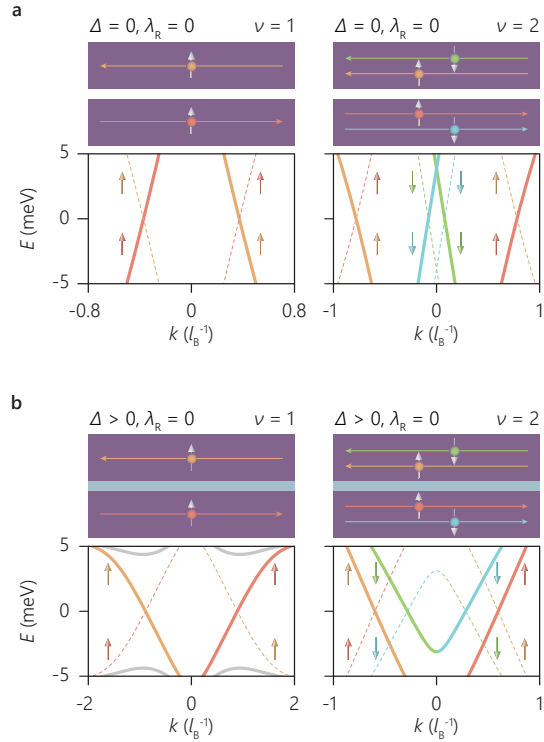

**FIG. 7. Evolution of Bogoliubov-de Gennes spectrum when including superconductivity without spin-orbit coupling. a,** Illustration of the edges separated by vacuum for $\nu=1, 2$, and their calculated Bogoliubov-de Gennes spectrum. Momentum $k$ in units of $l_B^{-1}$ with $l_B$ the magnetic length. $k$ to $-k$ symmetry is a result of physical ($Z_2$) symmetry—a reflection with respect to the separating vacuum (or superconductor for **b**), which exchanges the left-mover and the right-mover. Spectrum is doubled to show both electron-like (solid lines) and hole-like (dashed lines) excitations. Color code indicates the spin and the direction of propagation. **b,** Inclusion of superconductivity ($\Delta>0$) without spin-orbit coupling ($\lambda_R=0$) does not affect the spin-polarization. This leaves the zero-energy crossings spin polarized which cannot be gapped by an s-wave superconductor—an induced pairing gap $\Delta_{ind}$ does not open. **b** same as Figure 2c. For these simulations and the ones presented in Figure 2, the system parameters are set as follows: $\Delta_1= 0.3\varepsilon_0$, $\Delta_2= 0.2\varepsilon_0$ ($\Delta_{1,2}= 0$ for vacuum), $m_s = 3\varepsilon_0$, $m_n = 0.06\varepsilon_0$, $\mu_s = 8\varepsilon_0$ ($\mu_s = 0$ for vacuum), $\mu_n = 0.2\varepsilon_0$ (for $\nu = 1$), $\mu_n = 0.45\varepsilon_0$ (for $\nu = 2$), $g_n = 0.2\varepsilon_0$, $\lambda_{Ry} = \lambda_{SO} = 0$ and $\lambda_{Rx} = 0.2\varepsilon_0$ ($\lambda_{Rx} = 0$ for no spin-orbit), $W_s = 6l_B = 1.7\xi_0 = 62.7$ nm, $\varepsilon_0 = 89$ meV.

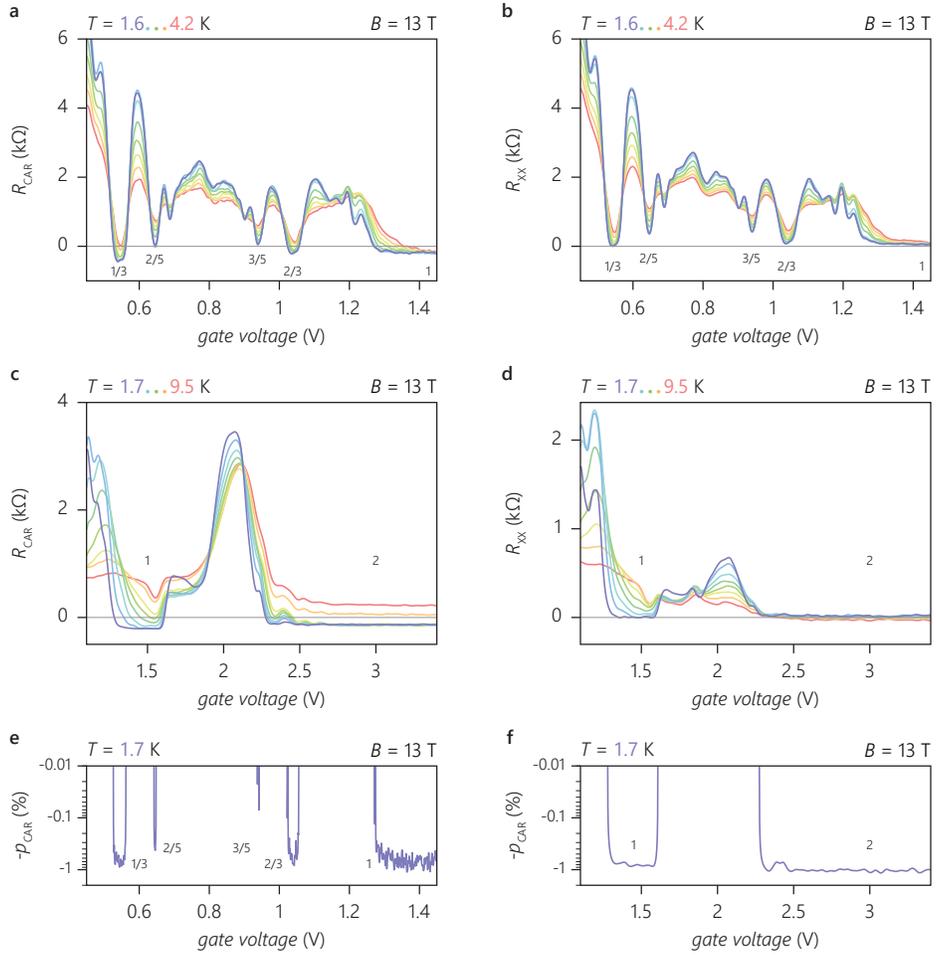

FIG. 8. Crossed Andreev reflection in fqH in Device 2. **a, b,** $R_{CAR}$ and $R_{XX}$ as a function of gate voltage at $B=13$ T for different temperatures $T$. $R_{CAR}<0$ at the highlighted fillings indicates crossed Andreev reflection (CAR). **c, d,** same as **a** and **b** but for integer fillings 1 and 2 in a larger $T$ range. **e, f,** $p_{CAR}$ at $T=1.7$ K respectively from **a** and **c**.

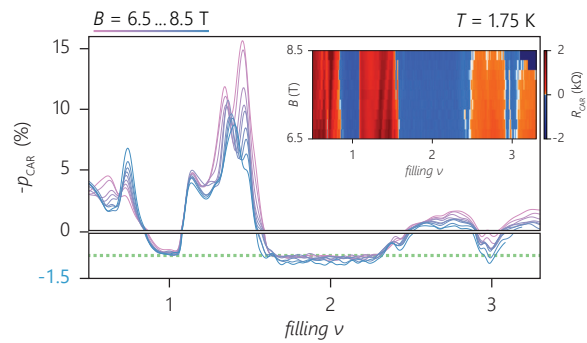

**FIG. 9. Spin-orbit coupling in Device 2. a,** $p_{CAR}$ as a function of filling $v$. Integer fillings, including the spin-polarized $v=1$, have a comparable $p_{CAR}$, an evidence for the pairing mechanism enabled by spin-orbit coupling. The dashed line is the $p_{CAR}$ expected for $Z\sim0.45$. Inset shows $R_{CAR}$ measured at $B$=6.5...8.5 T. The filling axis has been adjusted such that the measured fillings align with the integer values of the axis. This procedure to convert gate voltage to filling is present in this Figure and in Figure 2a and b.

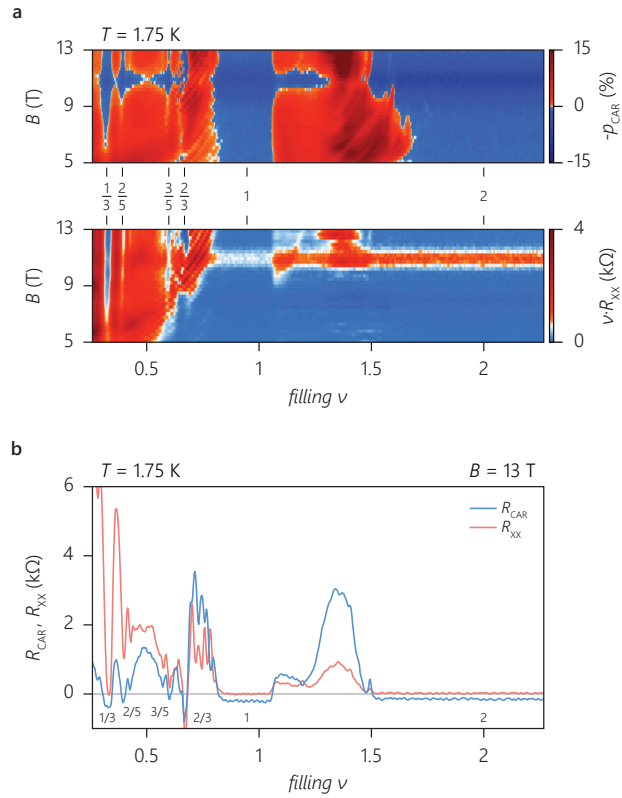

**FIG. 10. Magnetic field dependence in Device 2. a,** $p_{CAR}$ and $R_{XX}$ (normalized for different $\nu$) as a function of filling and magnetic field $B$. Crossed Andreev reflection (CAR, $R_{CAR}<0$) is seen for all well-developed $\nu$. Bulk conduction suppresses CAR observed for lower values of $B$ or for fractional $\nu$ with smaller excitation gaps. **b,** $R_{CAR}$ and $R_{XX}$ at $B=13$ T from **a**. Bulk conductance ($R_{XX}$) is negligible for all highlighted fillings except $\nu=2/3$, a complication related to equilibration in the contact, which is limited to this measurement.

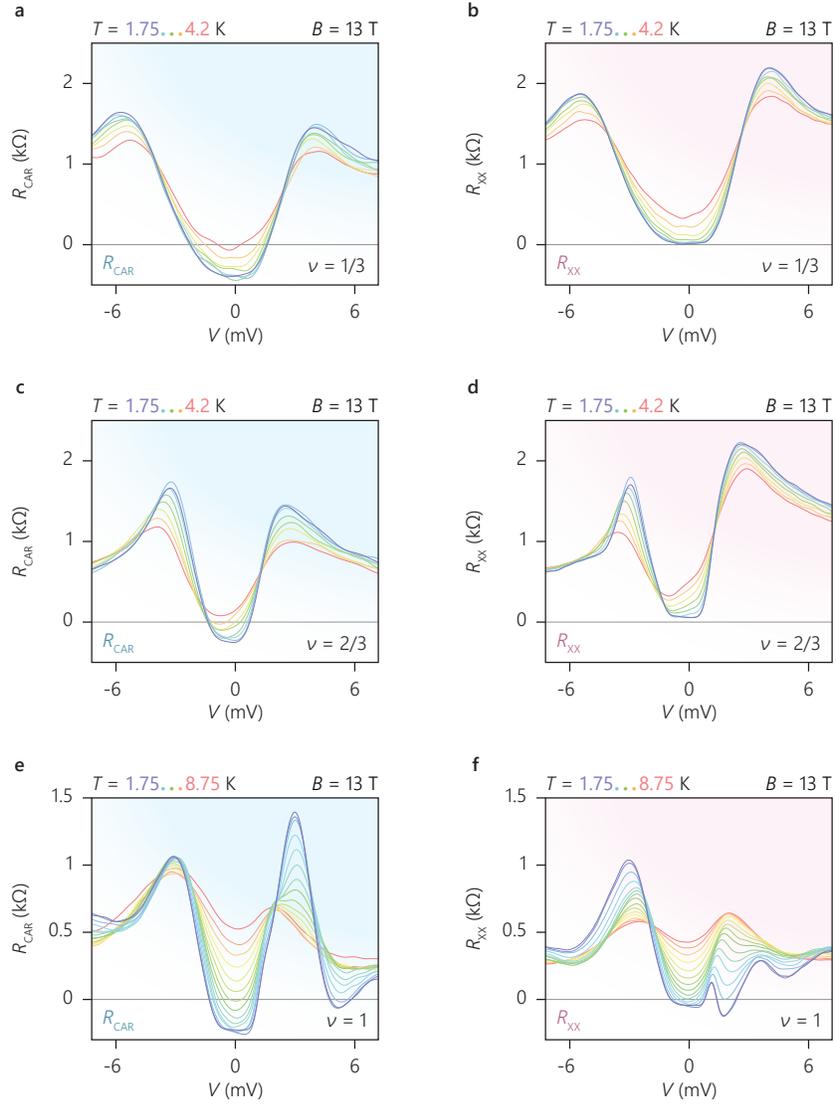

FIG. 11. **Transport spectroscopy in Device 2 (part 1).** a-f, $R_{CAR}$ and $R_{XX}$ at several fillings as a function of the incoming edge mode potential $V$ (excitation) for different temperatures $T$. Crossed Andreev reflection (CAR, $R_{CAR}<0$) is limited to below $|eV|\sim 1$ meV. Increasing excitation and $T$ suppress CAR by increasing bulk conductance for these fillings in this device.

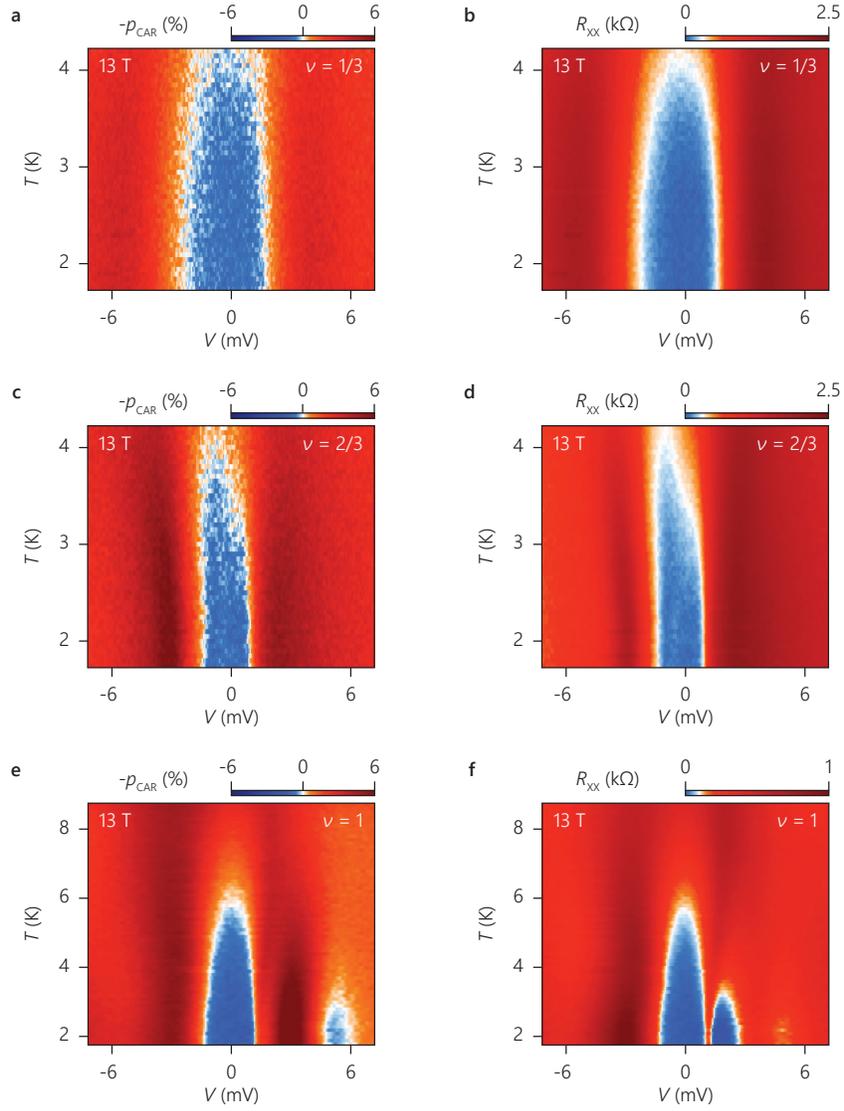

**FIG. 12. Transport spectroscopy in Device 2 (part 2). a-f,** $p_{CAR}$ and $R_{XX}$ at several fillings as a function of the incoming edge mode potential $V$ (excitation) for different temperatures $T$. Crossed Andreev reflection (CAR, $R_{CAR}<0$) is limited to below $|eV|$~1 meV, $T$~4 K for $\nu$=1/3 and 2/3, and $T$~6 K for $\nu$=1.

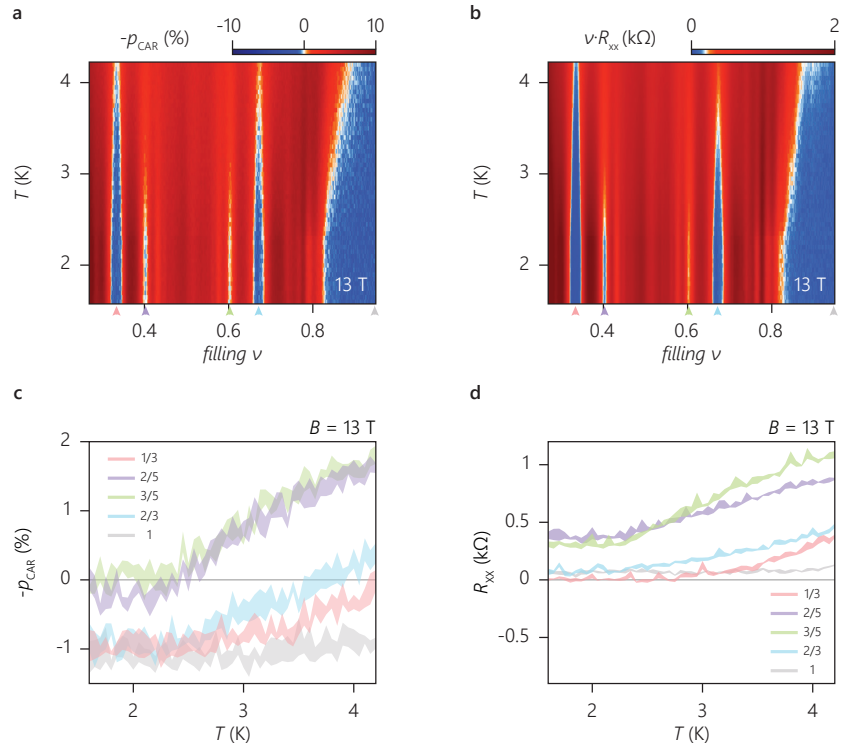

**FIG. 13. Temperature dependence in Device 2 (part 1, fractional fillings). a, b,** $p_{CAR}$ and $R_{xx}$ (normalized for different $\nu$) as a function of filling for varying $T$. Crossed Andreev reflection ($p_{CAR} > 0$) is seen for the well-developed fillings $\nu = 1/3$, $2/3$, and 1, as well as the fillings with smaller excitation gaps $\nu = 2/5$ and $3/5$, all highlighted with arrows. **c, d,** Vertical cuts respectively from **a** and **b**. The shades represent the standard deviation.

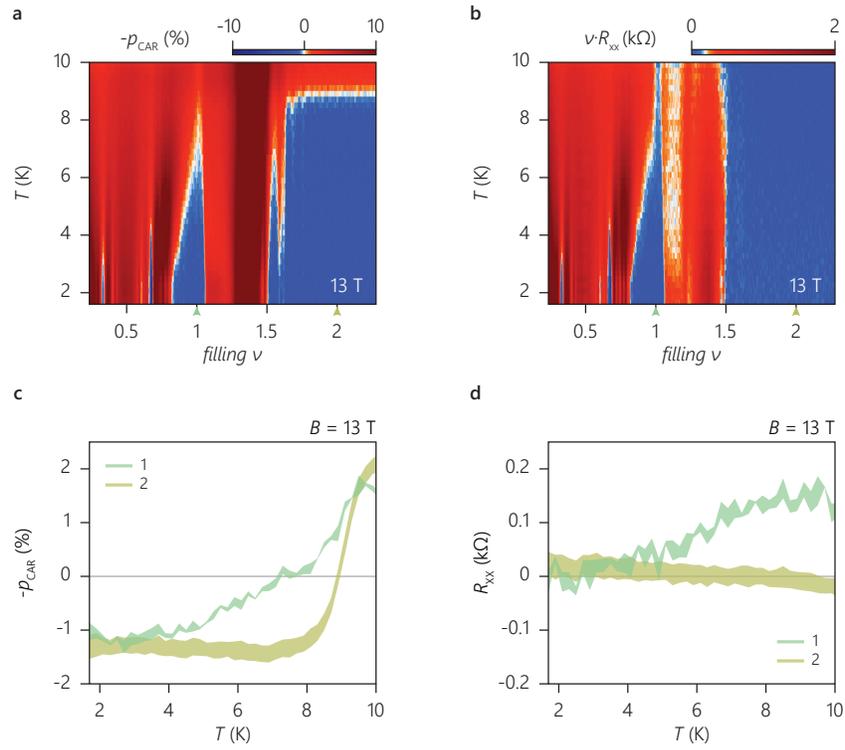

**FIG. 14. Temperature dependence in Device 2 (part 2, integer fillings). a, b,** $p_{CAR}$ and $R_{xx}$ (normalized for different $\nu$) as a function of filling for varying $T$. Crossed Andreev reflection ($p_{CAR}>0$) is seen for both integer fillings $\nu=1$ and 2. **c, d,** Vertical cuts respectively from **a** and **b**. The shades represent the standard deviation. Crossed Andreev reflection at $\nu=2$ is undisturbed by bulk conductance and suppressed when the NbN superconductor turns normal at $T\sim9$ K.

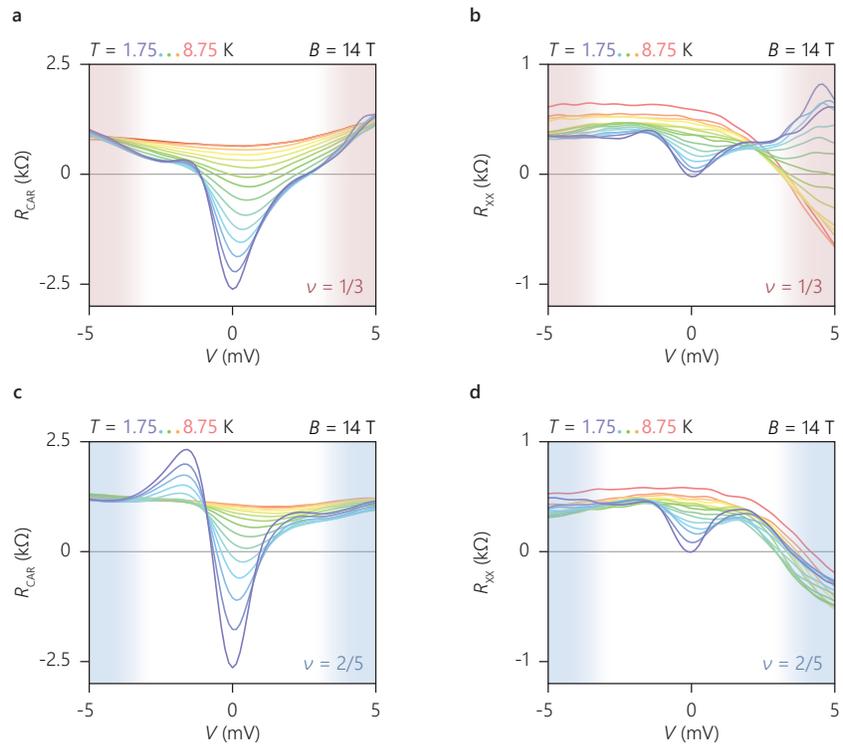

FIG. 15. **Transport spectroscopy in Device 1 showing $R_{CAR}$ and $R_{XX}$. a, b,** $R_{CAR}$ and $R_{XX}$ for $v=1/3$ as a function of the incoming edge mode potential $V$ (excitation) for different temperatures $T$. Crossed Andreev reflection (CAR, $R_{CAR}<0$) is limited to below $|eV|\sim 1$ meV. **c, d,** Same as **a** and **b** but for $v=2/5$ which shows a $V$ and $T$ dependence similar to that for $v=1/3$.

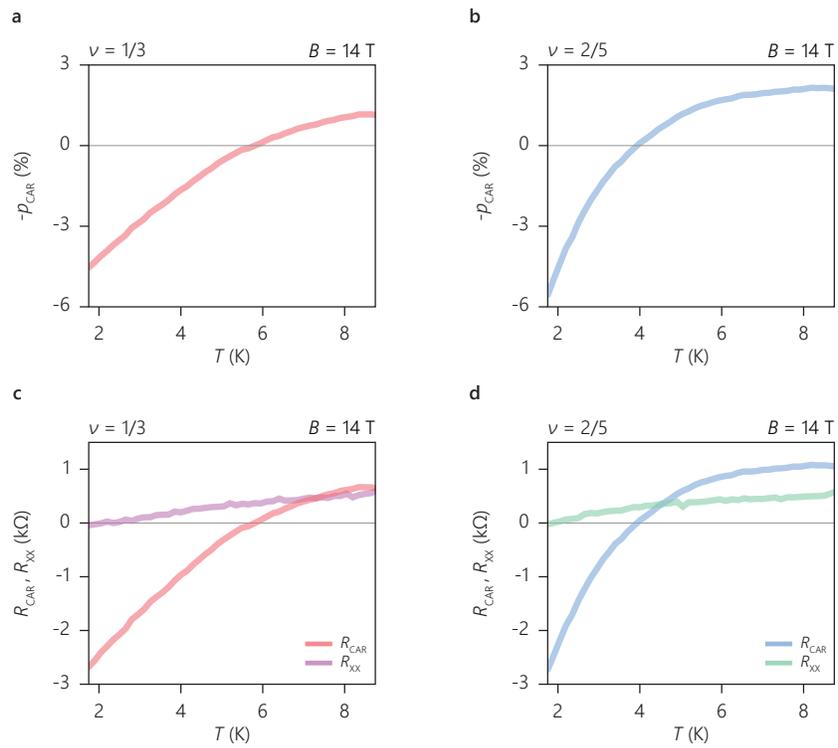

**FIG. 16. Temperature dependence of crossed Andreev reflection at $\nu=1/3$ and $2/5$ in Device 1.** Probability of crossed Andreev reflection in fillings $\nu=1/3$ and $2/5$ is rapidly increasing with decreasing $T$ down to the lowest temperature, at which bulk conductance vanishes ($R_{XX}\sim 0$). Data extracted from Figure 5a-d.

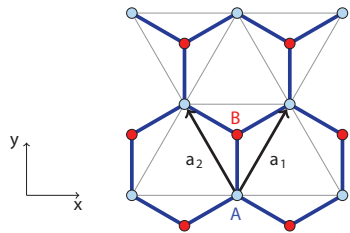

**FIG. 17. Honeycomb lattice of carbon atoms in graphene.** Armchair and zigzag boundaries are along $y$ and $x$ directions, respectively. Here, we have $\mathbf{a}_1 = a\left(\frac{1}{2}\hat{\mathbf{x}} + \frac{\sqrt{3}}{2}\hat{\mathbf{y}}\right)$, and $\mathbf{a}_2 = a\left(-\frac{1}{2}\hat{\mathbf{x}} + \frac{\sqrt{3}}{2}\hat{\mathbf{y}}\right)$ where $a$ is the lattice constant.

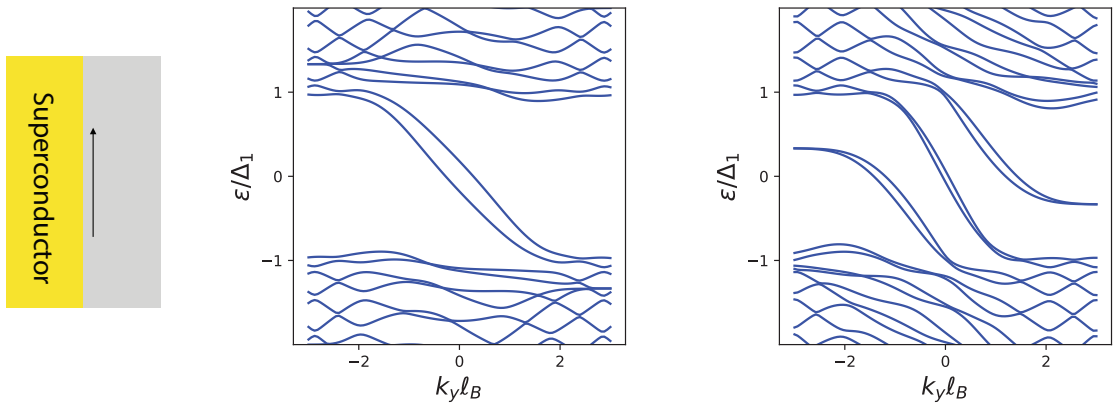

**FIG. 18. Graphene armchair edge modes near a superconductor.** Plotted for the lowest two quantum Hall states $\nu = 2$ and $\nu = 6$ (this is an analog of calculations in ref. [53] for graphene). Here, we set $\Delta_1 = 0.3\varepsilon_0$, $\Delta_2 = 0$, $m_s = 3\varepsilon_0$, $\mu_s = 8\varepsilon_0$, $m_n = 0$. We set $\mu_n = 0.4\varepsilon_0$ (**left**) and $1.1\varepsilon_0$ (**right**).

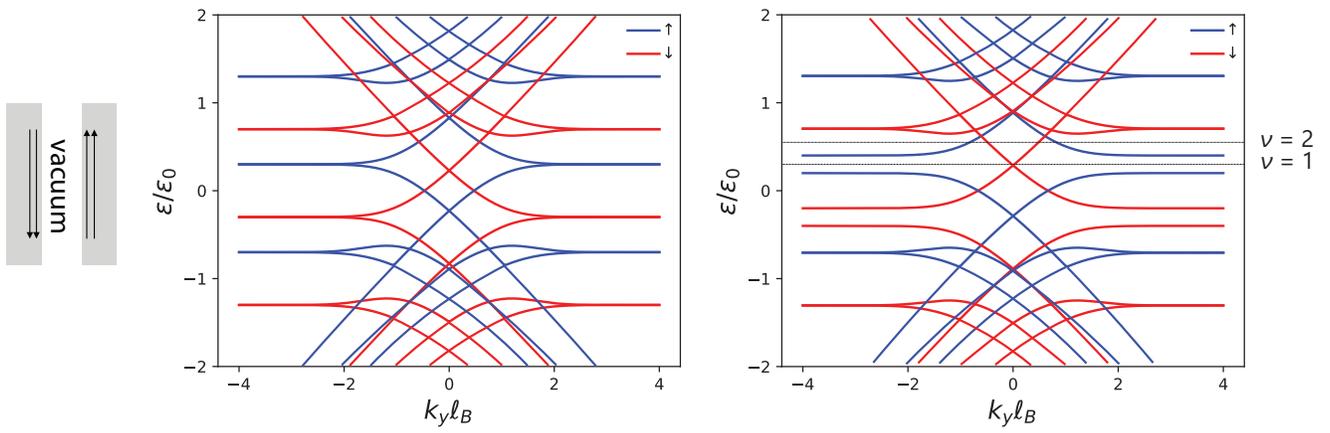

**FIG. 19. Edge modes of two half-infinite graphene quantum Hall systems separated by an insulator. Left** is with, **right** is without inversion breaking mass term. We set the parameters for the intermediate insulating region as $m_s = 3\varepsilon_0$, $\mu_s = 0$. The width of the insulating region is $W_s = 8l_B$. For the graphene regions $\mu_n = 0$, $g = 0.3\varepsilon_0$. We set $m_n = 0$ (**left**) and $m_n = 0.1\varepsilon_0$ (**right**).

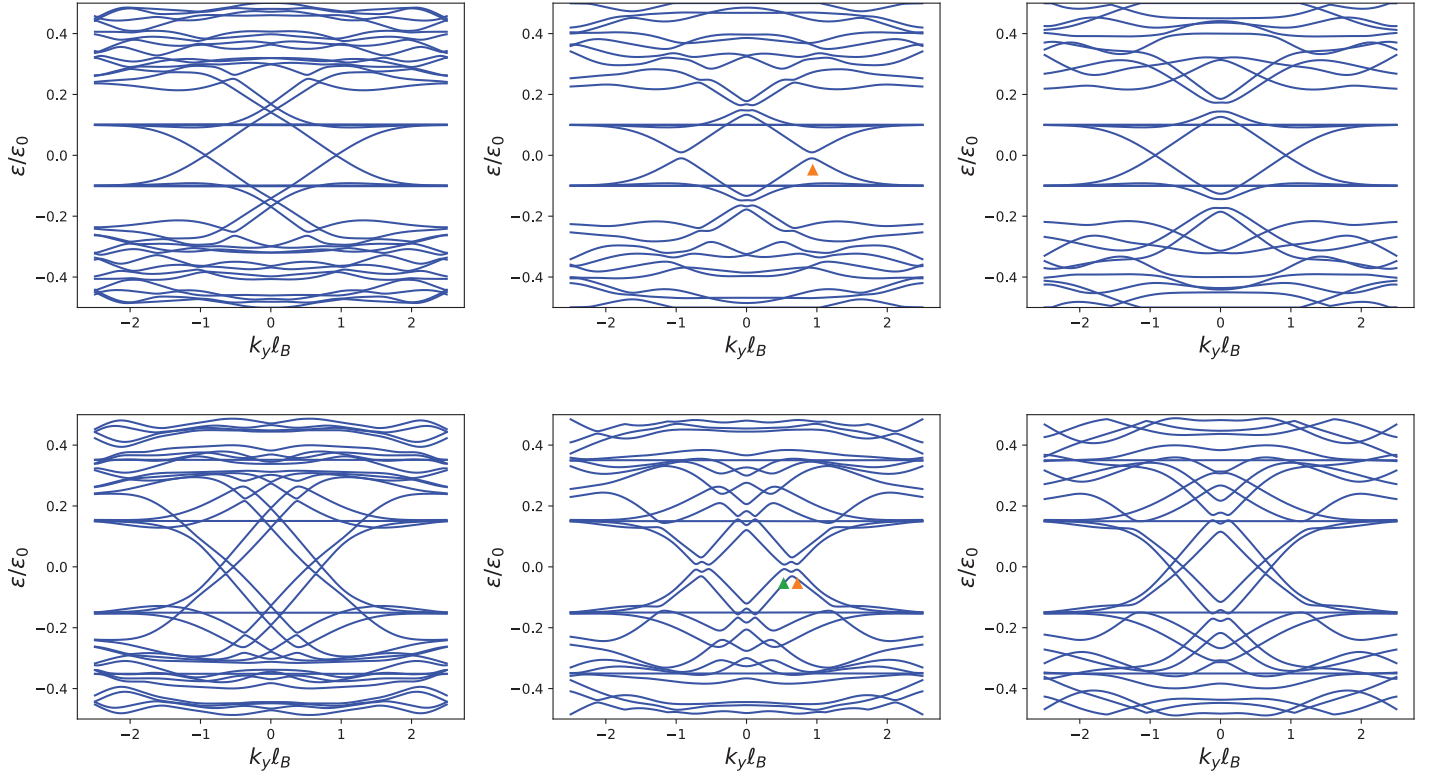

**FIG. 20. Bogoliubov spectrum.** (**Top row**) $\nu = 1$ with $\mu_n = 0.3\varepsilon_0$ and (**bottom row**) $\nu = 2$ with $\mu_n = 0.55\varepsilon_0$ (see the right panel of Figure 19 for the location of the chemical potential). **Left column** represents the case of a thick superconductor $W_s = 15 l_B = 10\xi_0$. **Middle and right column** correspond to a thin superconductor $W_s = 6 l_B = 4\xi_0$. There is no energy gap in the thick regime, while there is a gap opening in the thin regime when spin-orbit coupling is present (**middle column**). For reference, we provide **right column** which has no spin-orbit coupling. In the other panels we set $\lambda_{Rx} = \varepsilon_0$. It is evident that either turning off the spin-orbit coupling or making the superconductor thick prevent the edge modes from hybridizing and lead to a gapless spectrum with propagating Andreev edge states along the qH–superconductor interface. For these simulations, the system parameters are set as follows: $\Delta_1 = 0.5\varepsilon_0$, $\Delta_2 = 0.6\varepsilon_0$, $m_s = 3\varepsilon_0$, $\mu_s = 8\varepsilon_0$, $\lambda_{Ry} = \lambda_{SO} = 0$, $\lambda_{Rx} = \varepsilon_0$.

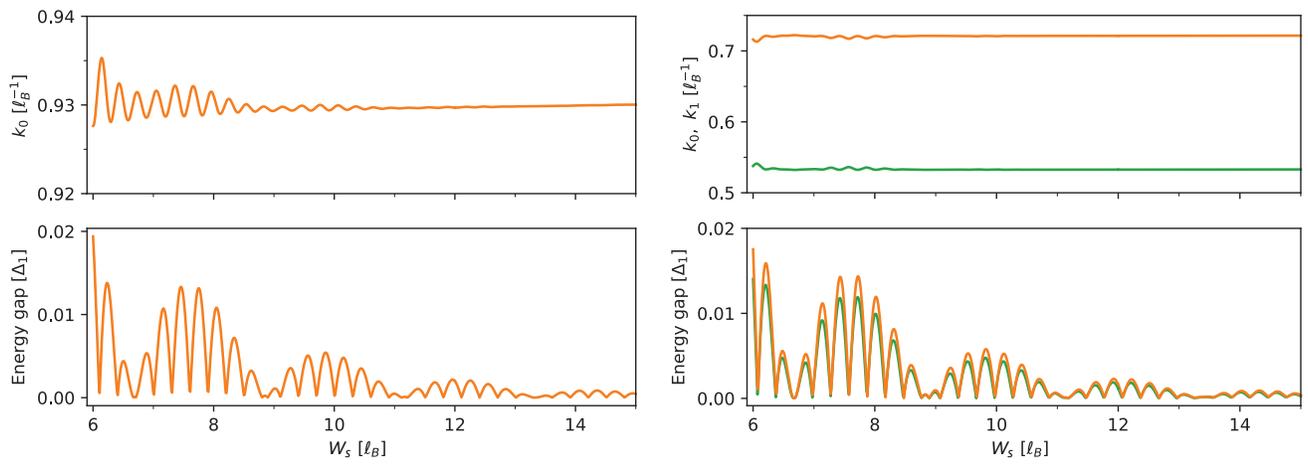

**FIG. 21. Superconductor thickness dependence.** $k$ of BdG energy minima $\pm k_0$, and $\pm k_1$ (only for $\nu = 2$), marked in the middle column of Figure 20, and their corresponding energy gap. As the thickness of the superconductor is increased, the gap in the BdG spectrum decreases. Other parameters are given in Figure 20.

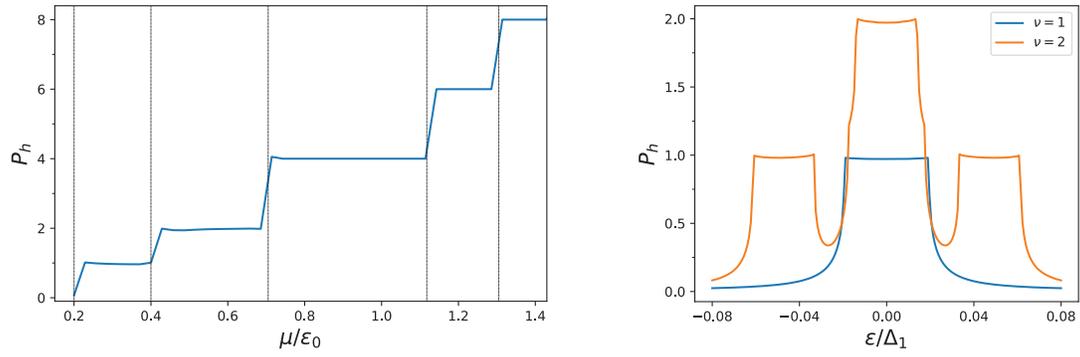

FIG. 22. **Collective probability of crossed Andreev reflection.** Plotted as a function of chemical potential (**left**), and bias voltage (**right**). Here, $W_s = 6l_B = 4\xi_0$. The rest of the parameters are given in Figure 20.

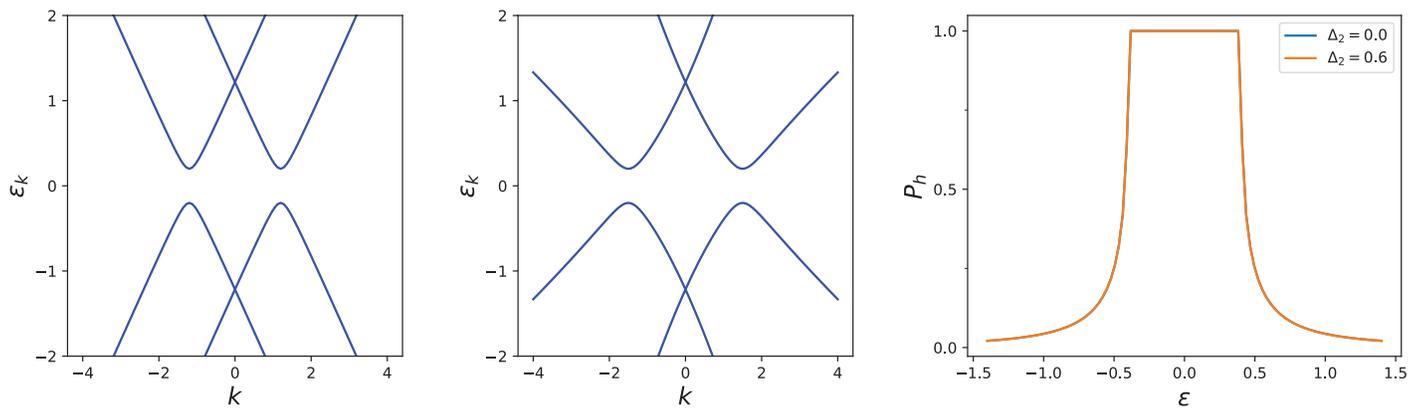

**FIG. 23. Bogoliubov spectrum at $\nu = 1$ edge from the effective theory.** $k_0 = 1.2$, $\Delta_1 = 0.4$. We set $\Delta = 0$ (**left**) and $0.6v_F$ (**middle**). **Right** is the collective probability of crossed Andreev reflection for the same parameters.

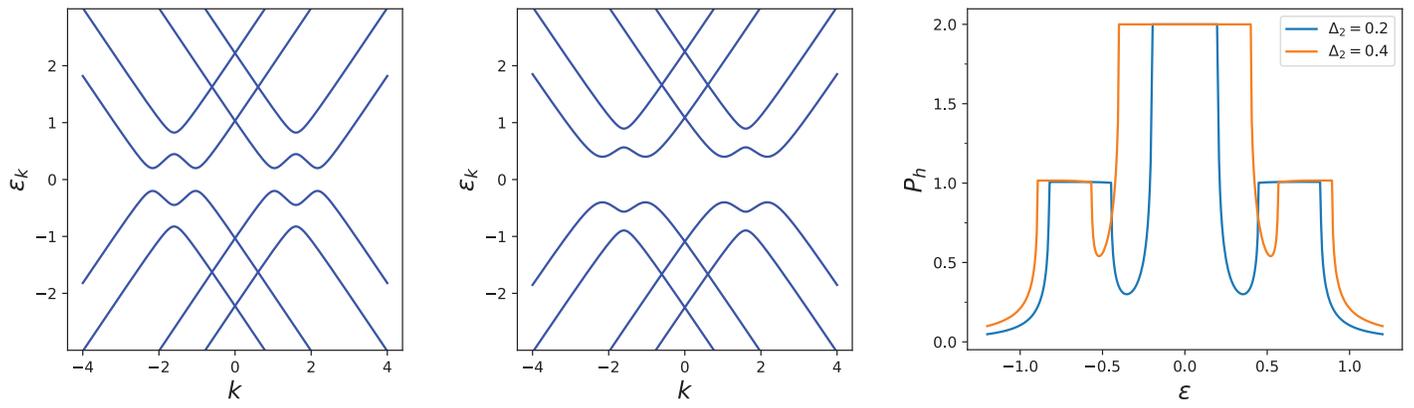

**FIG. 24. Bogoliubov spectrum at $\nu = 2$ edge from the effective theory.** $k_{F\uparrow} = 2.2$, $k_{F\downarrow} = 1$, $\Delta_1 = 0.2$, $\widetilde{\Delta}_1 = 0$. We set $\Delta_2 = 0.2$ (**left**) and $0.4 v_F$ (**middle**). **Right** is the collective probability of crossed Andreev reflection for the same parameters.

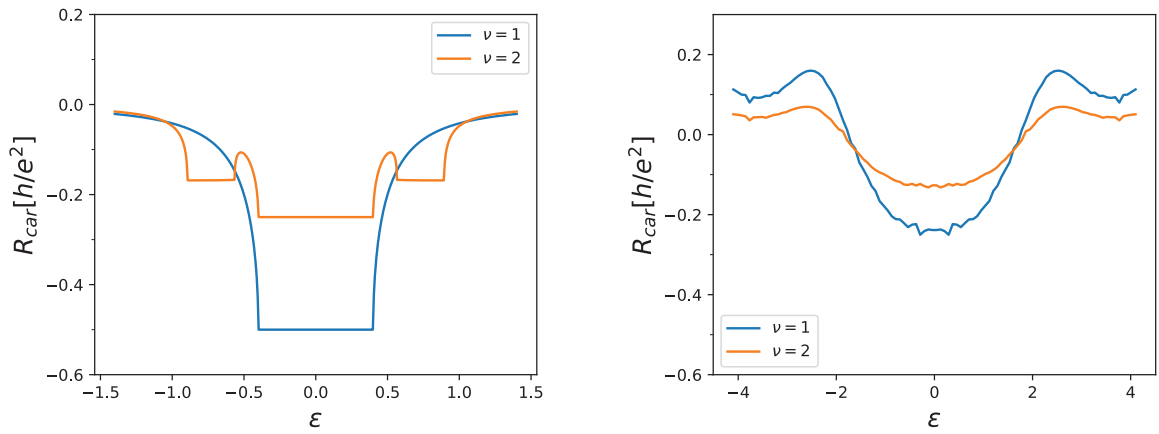

FIG. 25. Comparison of $R_{CAR}$ for $\nu = 1$ and $2$. **Left** is the ideal limit, **right** is the disordered pairing with $Z = 0.4$.

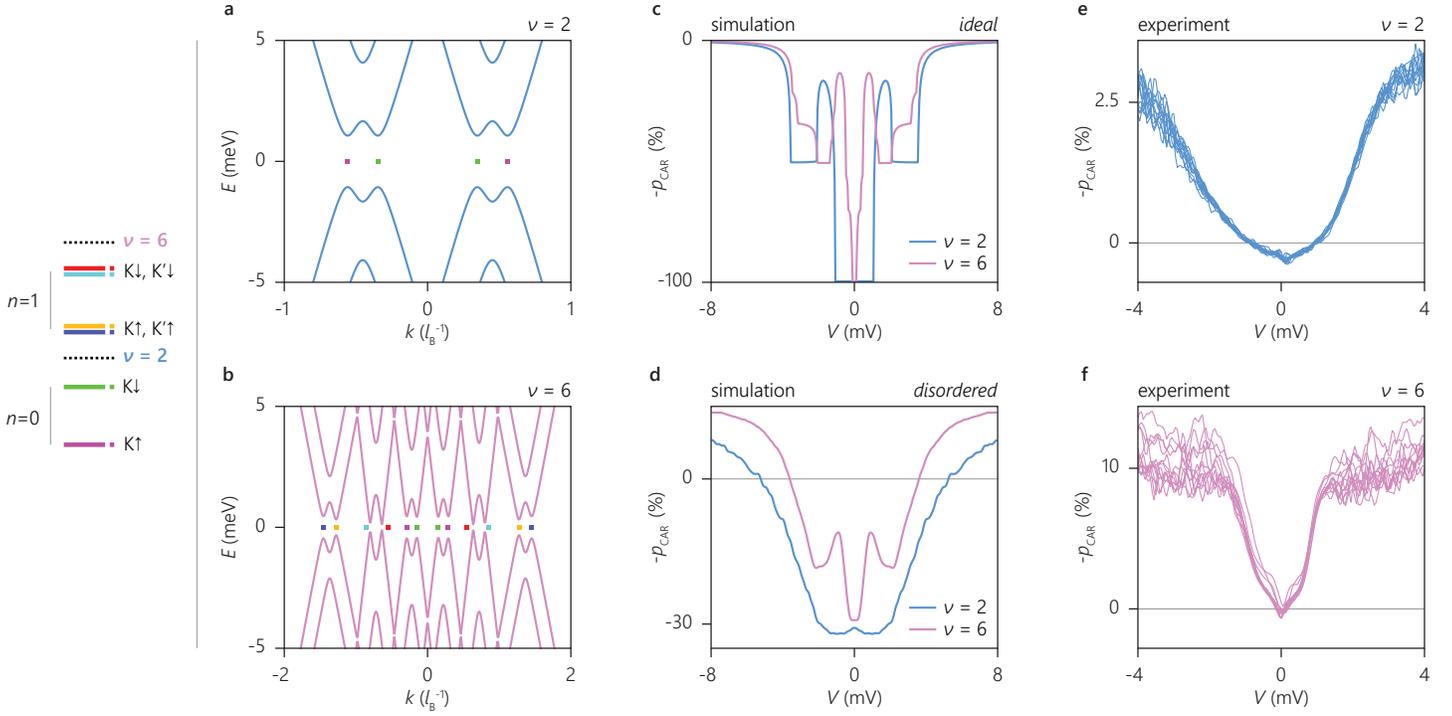

**FIG. 26. Bogoliubov-de Gennes spectrum, calculated CAR response, and transport spectroscopy in Device 1 for $\nu=2$ and 6. a, b,** Bogoliubov-de Gennes spectrum of $\nu=2$ and 6 calculated using 2D simulations. Hybridization of the inner edge modes (those from lower Landau levels) is stronger due to their closer proximity to the superconductor, leading to larger gaps around $k$ for which band crossings occur. Coloured squares indicate the Landau levels whose edge modes hybridize. **c, d,** CAR response calculated using the effective edge theory in the ideal and the disordered case. For the clean case, CAR asymptotically vanishes at high energies. Including disorder limits the energy range for which $p_{CAR}>0$. This implies that the pairing gap and disorder together determine the energy range of CAR in a realistic setup. The parameters used in the effective theory are extracted from the 2D simulations for which the system parameters are the same as in Figure 7 ($\mu_n = 1.2\varepsilon_0$ for $\nu = 6$). **e, f,** $p_{CAR}$ at $\nu=2$ and 6 as a function of the incoming edge mode potential $V$ (excitation) for different gate voltages spanning the entire qH plateau region. CAR is limited to below $|eV|\sim1$ meV for $\nu=2$ and below $|eV|\sim0.2$ meV for $\nu=6$. **a** same as in Figure 2d. **e** and **f** same as Figure 2e and f.

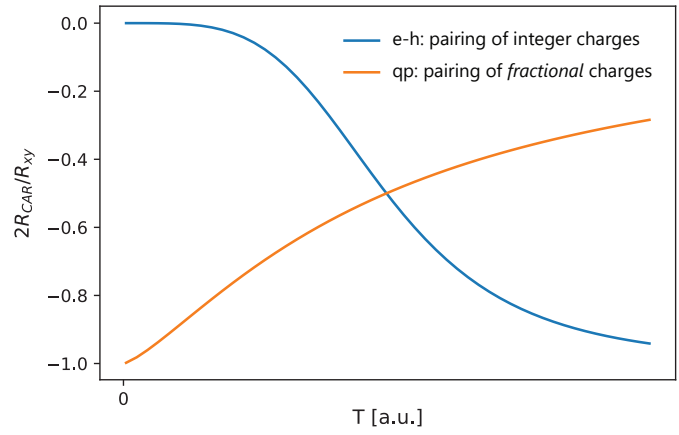

**FIG. 27. Temperature dependence of $R_{CAR}$ in a fqH edge mode.** $R_{CAR}$ at the edge of Laughlin states shows different temperature dependences for two separate pairing scenarios. In one scenario, pairing occurs between fractional charges, that is, crossed Andreev reflection converts fractionally charged quasiparticles to fractionally charged quasiholes. This mechanism results in an increasing $|R_{CAR}|$ with decreasing temperature $T$. Such $T$ dependence is not present in the alternative scenario where pairing occurs between integer charges which are formed by bunched fractional charges. Here, crossed Andreev reflection converts particles with integer charge to holes with integer charge. In this case, $R_{CAR}$ vanishes at $T=0$ because in fqH, integer charges are composed of excitations which are nonexistent at zero $T$. The probability to form integer charges is then proportional to a large power of $T$. Both these scenarios are specific to fqH. No temperature dependence is expected for integer quantum Hall edge modes. CAR of fractional charges (first scenario) implies addition of fractional charges to a fractional topological superconductor hosting parafermions—a mechanism similar to Majoranas allowing for transport of single charges in a topological superconductor, which is not possible in a conventional superconductor. A similar mechanism analogous to CAR of fractional charges occurs in a quantum point contact (QPC) separating two fqH regions. Here, fractional charges first tunnel through the QPC[62,63], and then are drained from a metal electrode which does not host fractional excitations. In this analogy, the tunneling of fractional charges through a QPC is similar to our CAR of fractional charges. The metal electrode draining the edge modes after a QPC is similar to our conventional superconductor away from the proximitized region.

| Device ID | Completed fabrication steps | Completion date | Device ID | Completed fabrication steps | Completion date |
|---|---|---|---|---|---|
| SOY1 | Heterostructure assembly, alignment markers, defining top graphite. | ---- | SOY12 | Heterostructure assembly, alignment markers. | ---- |
| SOY2 | Heterostructure assembly, alignment markers, defining top graphite. | ---- | SOY13 | Heterostructure assembly, alignment markers, defining top graphite, defining heterostructure, normal contacts, superconducting contact. | 17 Apr 2019 |
| SOY3 | Heterostructure assembly, alignment markers, defining top graphite, normal contacts. | ---- | SOY14 | Heterostructure assembly, alignment markers, defining top graphite, defining heterostructure. | ---- |
| SOY4 | Heterostructure assembly, alignment markers. | ---- | SOY15 | Heterostructure assembly, alignment markers, defining top graphite, defining heterostructure, normal contacts, superconducting contact, top graphite connection bridges. | 31 May 2019 |
| SOY5 | Heterostructure assembly, alignment markers, defining top graphite, normal contacts, superconducting contact, defining heterostructure. | 12 Oct 2018 | SOY16 | Heterostructure assembly. | ---- |
| SOY6 | Heterostructure assembly, alignment markers, defining top graphite, normal contacts, superconducting contact, defining heterostructure. | 25 Oct 2018 | SOY17 | Heterostructure assembly, alignment markers, defining top graphite, defining heterostructure, normal contacts, superconducting contact, top graphite connection bridges, defining top graphite local gates. | 26 Jul 2019 |
| SOY7 (**Device 1**) | Heterostructure assembly, alignment markers, defining top graphite, normal contacts, superconducting contact, defining heterostructure. | 11 Jan 2019 | SOY18 (**Device 2**) | Heterostructure assembly, alignment markers, defining top graphite, defining heterostructure, normal contacts, superconducting contact, top graphite connection bridges, defining top graphite local gates. | 12 Aug 2019 |
| SOY8 | Heterostructure assembly, alignment markers. | ---- | SOY19 | Heterostructure assembly. | ---- |
| SOY10 | Heterostructure assembly, alignment markers, defining top graphite, normal contacts, top graphite connection bridges, superconducting contact, defining heterostructure. | 11 Mar 2019 | SOY20 | Heterostructure assembly, alignment markers, defining top graphite, device etch, normal contacts. | ---- |
| SOY11 | Heterostructure assembly, alignment markers. | ---- | SOY21 | Heterostructure assembly, alignment markers, defining top graphite, defining heterostructure, normal contacts, superconducting contact, top graphite connection bridges, defining top graphite local gates. | 7 Jul 2020 |

**Table 1. Fabricated devices.**

| Device ID | # of cooldowns | Result | $p_{CAR}$ at fractional $v$ |
|---|---|---|---|
| SOY5 | 2 | No evidence for Andreev reflection at $B=0$ T (no conductance enhancement). | |
| SOY6 | 2 | No evidence for Andreev reflection at $B=0$ T (no conductance enhancement). | |
| SOY7 (**Device 1**) | 4 | Reported data from two different cooldowns. | $p_{CAR}$ at $v=1/3$, 2/5, 4/3 stronger than that at integer $v$ (Figure 1-3, 5 and Extended Data Figure 10, 11) |
| SOY10 | 2 | Andreev reflection (conductance enhancement) at $B=0$ T and high $B$ is present. Negative signal is present at high $B$ for integer and fractional states. No strong evidence in bias/temperature dependence or support from $R_{XX}$ to attribute the negative signal to CAR. | |
| SOY13 | 2 | No evidence for Andreev reflection at $B=0$ T (no conductance enhancement). | |
| SOY15 | 2 | Negative signal is present at high $B$ for integer and fractional states. Bias/temperature dependence is consistent with CAR. | $p_{CAR}$ at fractional $v$ comparable to that at integer $v$ (~1 % at $v=2$) |
| SOY17 | 1 | Low mobility device. | |
| SOY18 (**Device 2**) | 4 | Reported data from two different cooldowns. | Cooldown A: $p_{CAR}$ at $v=1/3$ stronger than that at integer $v$ (Figure 4) Cooldown B: $p_{CAR}$ at fractional $v$ comparable to that at integer $v$ (Extended Data Figure 3-9) |
| SOY21 | 1 | Global back gate leak prevents measurement. | |

**Table 2. Measured devices.**